\documentclass[11pt,a4paper]{article}
\usepackage{graphicx} 
\usepackage[margin=2.5cm]{geometry}
\usepackage{xcolor}
\usepackage{amsmath}
\usepackage{booktabs}
\usepackage{dcolumn}

\usepackage[backend=biber,style=nature,isbn=false,url=false,eprint=false,giveninits=true,maxcitenames=1,maxnames=5,uniquename=init]{biblatex}

\graphicspath{{./figures/}}

\AtEveryBibitem{
  \clearfield{note}
  \clearfield{url}
  \clearfield{issn}
  \clearfield{urldate}
  \clearfield{month}
  \clearfield{language}
}

\definecolor{bleucite}{RGB}{34,111,212}
\usepackage{hyperref}
\hypersetup{hidelinks}

\usepackage{setspace}
\usepackage{pdflscape}
\usepackage[defaultlines=2,all]{nowidow}

\usepackage{caption}

\usepackage{multirow}

\title{Fragmentation of a longitudinal population-scale social network: 
Decreasing
structural social cohesion in the Netherlands}
\author{Eszter Bokányi$^{1}$, Yuliia Kazmina$^2$, Eelke M. Heemskerk$^2$, Frank W. Takes$^{1,*}$}
\date{\vspace{-0.5cm}{\footnotesize
    $^1$Leiden University\\
    $^2$University of Amsterdam\\
    $^*$\href{mailto:f.w.takes@liacs.leidenuniv.nl}{f.w.takes@liacs.leidenuniv.nl}
}}

\begin{document}

\maketitle

\onehalfspacing

\begin{abstract}
Population-level dynamics of social cohesion and its underlying mechanisms remain difficult to study.
In this paper, we propose a network approach to measure the evolution of social cohesion at the population scale and identify mechanisms driving the change.
We use twelve annual snapshots (2010--2021) of a population-scale social network from the Netherlands linking all residents through family, household, work, school, and neighbor relations. Results show that over this period, social cohesion, quantified as average closure in the network, declines by more than 15\%. We demonstrate that the decline is not due to changes in demographic composition, but to rewiring in individual ego networks. Statistical models confirm a decreasing overlap of social contexts and greater geographical mobility as drivers. Residential relocation, however, temporarily increases closure, suggesting that local cohesion-seeking behavior can yield global network fragmentation, with implications for policies related to housing, urban planning, and social integration. 

\end{abstract}

\onehalfspacing

\section{Introduction}

Understanding the foundations of social cohesion that binds people into communities has long been a central concern of the social sciences \parencite{toennies1887gemeinschaft, durkheim1893division, wellman1979community}. 
A higher level of social cohesion reduces neighborhood violence through  mutual trust and the willingness to intervene for the common good  \parencite{sampson1997neighborhoods}, it is linked to increased trust and civic cooperation \parencite{putnam2000bowling}, as well as to lower suicide rates \parencite{durkheim2005suicide}, juvenile delinquency \parencite{shaw1942juvenile}, and better health outcomes at the individual level \parencite{subramanian2002social, poortinga2007perceptions}. 
Some empirical evidence suggests an overall decline in the sense of community and cohesion \parencite{putnam2000bowling}, together with a spreading ``loneliness epidemic'' \parencite{holt-lunstad2015loneliness}. Others argue that instead of a decline, contemporary communities restructure from being based on local groups, such as neighbors and family members living nearby, to consisting of geographically far more dispersed social connections \parencite{wellman1999networks}. All in all, results across different countries and at different timescales are mixed \parencite{larsen2013rise}. For instance, in the Netherlands, survey evidence indicates that declines in trust or participation are not clearly detectable \parencite{schmeets2014declining}. However, a key gap remains: large-scale longitudinal measurements that can decisively assess whether social cohesion is stable or changing and that provide an understanding about its general driving forces.

Social cohesion can be viewed as a structural phenomenon \parencite{simmel1964conflict}, captured by the extent to which an individual's social connections are also related to each other or participating in intersecting social circles \parencite{vasquesfilho2020transitivity}, leading to the formation of closed triads \parencite{newman2010networks}. Triadic closure is the micro-level process that generates dense, cohesive local groups in social networks \parencite{bianconi2014triadic,asikainen2020cumulative,peixoto2022disentangling}. Moreover, it can be used as an empirical measure of social capital \parencite{bourdieu1986forms,kazmina2025can} facilitating trust, common norms, and shared expectations \parencite{coleman1988social}. The level of triadic closure in a network is influenced by multiple different processes. At the group level, demographic changes such as increasing migration and falling fertility rates leading to an aging society increase the relative share of groups having higher levels of closure (migrants and the elderly) due to their life circumstances. At the individual level, a decrease in the overlap between different social contexts (e.g., work, neighborhood, associations) \parencite{feld1981focused,vasquesfilho2020transitivity} and increased short and long-term geographic mobility \parencite{festinger1950social, viry2012residential, dulmen2022places} both lead to more open, fragmented network structures.

Despite the longstanding interest in social cohesion, few studies have been able to empirically track how it evolves over time across a large population, let alone investigate both the group-level or individual-level drivers of the change. Assessing the dynamics of social cohesion through closure requires longitudinal, population-scale social network data coupled with high-quality demographic information that was not accessible before.
Traditional social network surveys (e.g., \cite{fischer2018uc,volker2020social,share-eric2024survey-2}), while representative, are hard to scale up beyond a certain size, are costly to implement, and often contain only a few ties and little information about connections between the respondents' immediate network neighbors, which would be crucial for capturing triadic closure. Mobile communication records \parencite{onnela2007structure, eagle2009inferring, calabrese2011interplay, saramaki2014persistence} are often proprietary, and distorted by differing market shares and by communication preferences shifting to online platforms \parencite{wellmann2019are, khemakongkanonth2025empirical}. Online social networks, while able to uncover large-scale network structures \parencite{park2018strength}, suffer from node and edge sampling biases, often lack detailed demographic or location information on the nodes, as well as contextual information on the edges \parencite{bokanyi2023anatomy}. 

To overcome these limitations, recent work uses population-scale social networks of entire countries derived from central administrative registers~\parencite{bokanyi2023anatomy, cremers2025unveiling, panayiotou2025anatomy}. These networks are based on formal social relationships and were found to reflect the connectivity and community structure found in online social networks \parencite{menyhert2025connectivity}. Population-scale social network data has proven consequential for various societal outcomes, including educational attainment \parencite{garcia2025prediction}, COVID-19 spread \parencite{candogan2025network,hedde2024predicting,garcia-bernardo2024assessing}, segregation \parencite{kazmina2024socioeconomic}, perceptions on immigration \parencite{kazmina2024contact}, or upward economic mobility \parencite{kazmina2025can}. In this work, we leverage yearly population-scale social networks of the Netherlands between 2010 and 2021 \parencite{vanderlaan2022person}, which contain family, household, work, school, and neighbor ties. They also include official demographics and residential location for each individual, allowing us to investigate changes in social network closure alongside potential drivers. These latter include group-level demographic shifts and individual-level changes in the overlap of social contexts and geographic mobility. 

To begin with, we provide the first comprehensive longitudinal analysis of the structural evolution of a population-scale social network of Dutch residents over more than a decade. 
Most notably, we find that while the average number of social connections is almost unchanged, the average closure in the network declines by more than 15\%. 
Second, through the combination of a decomposition technique with ego network timeline clustering, we reveal that this change happens at the individual level, and is not due to group-level demographic shifts such as an aging society or increasing migration. 
Third, using a statistical model with two-way fixed-effect panel regressions, we test the underlying individual-level drivers changes in ego network closure. Here, we find that two major contributors: a decreasing share of multiplex edges capturing the extent of shared social contexts (for instance, family and school) and increasing geographical dispersion of network neighbors are driving closure down. Moreover, we observe what we coin a ``relocation paradox'': counterintuitively, relocation is associated with a short-term increase in closure, suggesting the creation of dense local ties post-move, despite the expectation that moving increases geographic dispersion of ties and as such, contributes to network fragmentation.  

By comprehensively mapping the evolution of the social network in a full population, our findings offer compelling evidence that contemporary societies are becoming more socially fragmented as captured by decreasing level of network closure. We confirm that this trend is not a group-level demographic change, but rather due to a decrease in the overlap of social contexts and increasing geographic mobility. Taken together, the decline in closure is consistent with a broader transformation towards a more mobile and digitally mediated society, in which improved transport and communication infrastructure and the growing ability to maintain relationships across larger distances reduces the structural importance of geographically co-located, overlapping social contexts. Yet, an open question is how this structural shift translates to lived and perceived social cohesion and whether policies or institutions, if any, can mediate its downstream consequences.
 
\section{Results}

The dataset we use for the investigation of longitudinal changes consists of yearly snapshots of the formal social connections in the Netherlands between 2010 and 2021, derived from country-level administrative registers \parencite{vanderlaan2022person}. The nodes in the yearly snapshots are all people registered in the Netherlands on 1st January of the given year. Connections represent people's social opportunity structures~\parencite{bokanyi2023anatomy}, covering family and household members, school classmates, work colleagues, and next-door neighbors. We refer to these connections as edges in different network layers. Demographic information and address location up to the neighborhood level are also available on all nodes from the administrative registers. For more details on data availability and the composition of this multilayer social network, see the \nameref{sec:mm} section. 

\subsection{Longitudinal population-scale network structure: Decreasing closure}

\begin{figure}[h!]
    \centering
    \includegraphics[width=\linewidth]{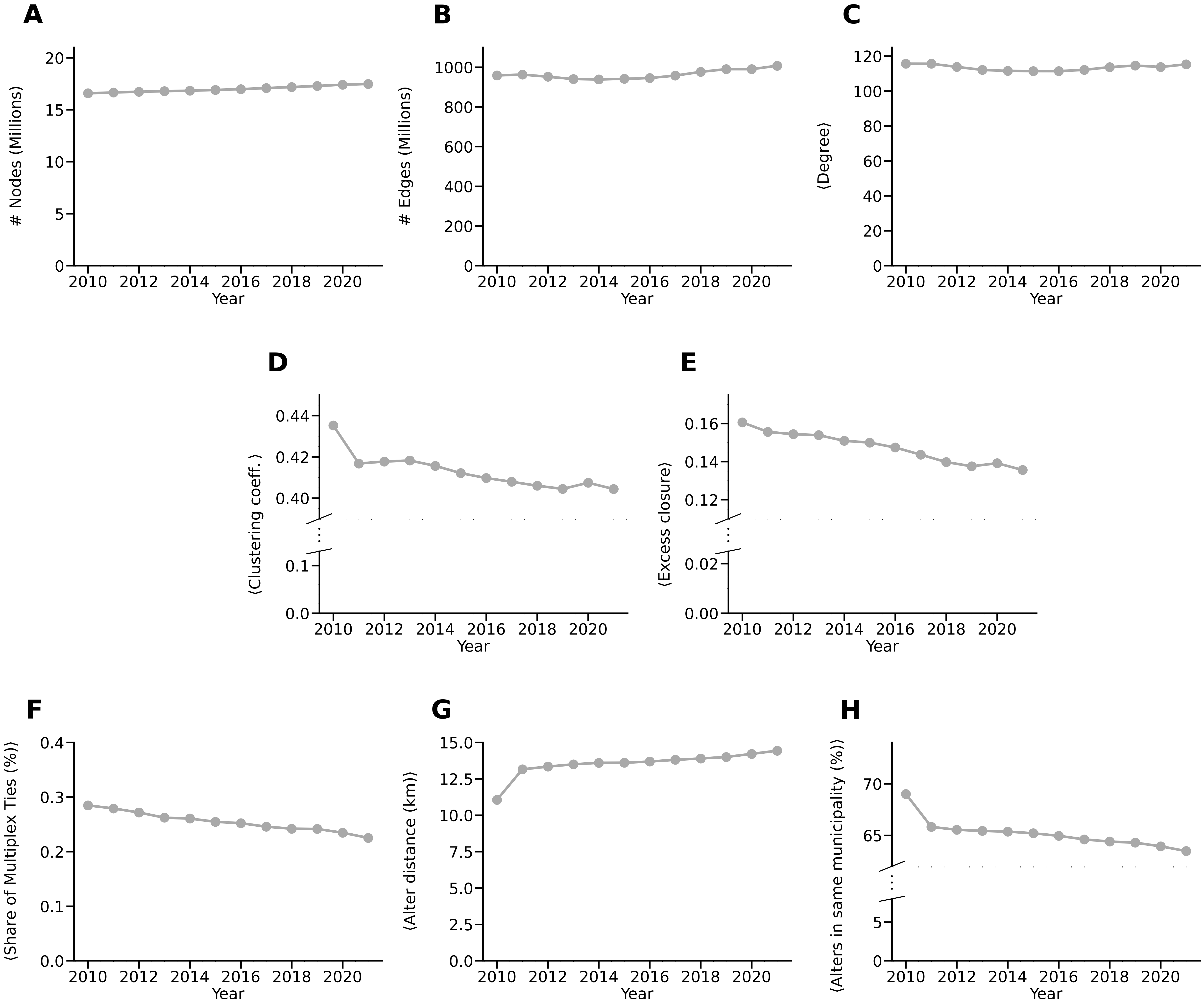}
    \caption{\textbf{Temporal evolution of the nationwide Dutch multilayer population-scale network, 2010--2021.} (A) Number of nodes. (B) Number of edges. (C)~Average degree. 
    (D) Average local clustering coefficient. (E) Average excess closure. (F) Share of multiplex ties.
    (G) Spatial dispersion in terms of the average alter distance (in km) and the share of alters in same municipality.}
    \label{fig:global_properties}
\end{figure}

First, to get an overview of longitudinal network properties, we look at the timeline of the total number of nodes and edges of the full yearly networks. In Figure~\ref{fig:global_properties}A, the \textit{number of nodes}, that is, the number of people registered in the country on the 1st of January each year, is steadily increasing over time, which reflects the country's positive net migration balance despite aging and falling fertility patterns. The \textit{number of edges} in Figure~\ref{fig:global_properties}B is also slightly increasing, leading to a relatively stable \textit{average degree} -- in other words, the average number of social connections for individuals, regardless of layer -- remains stable over time in Figure~\ref{fig:global_properties}C. 

Next, we look at two metrics to quantify the closure in 50k sampled ego networks of individuals: the average \textit{local clustering coefficient} (Figure~\ref{fig:global_properties}D), and \textit{average excess closure} (Figure~\ref{fig:global_properties}E). The sampling was done from the pool of all people who are present at least once in the yearly networks between 2010 and 2021 as further detailed in the  \nameref{sec:mm} section. The local clustering coefficient is calculated as the number of closed triangles in a node's ego network, without taking the different layers into account. In addition, we use excess closure, a metric designed specifically for multilayer register-based networks \parencite{bokanyi2023anatomy}, which is a normalized measure of triangles consisting of edges from more than one layers around egos. Excess closure discounts single-layer triads and captures local bridging and overlaps between different social contexts in multilayer networks where high-levels of within-layer clustering would mask these processes. Notably, we find that despite the stable average degree, both the average local clustering coefficient (Figure~\ref{fig:global_properties}C) and the average excess closure (Figure~\ref{fig:global_properties}D) show a \emph{decreasing trend}. The clustering coefficient drops around 7\% as compared to its initial levels, and the excess closure by as much as 15.5\% as compared to the initial levels over the observation timeframe. Exact va and values and descriptives of these yearly network metrics can be found in Tables~\ref{tab:global_properties}-\ref{tab:si_spatial_metrics}.

We also investigate the footprints of shared social contexts and geographic mobility in the ego networks. First, we measure multiplexity, the share of alters that are connected to an ego by edges in multiple different layers, such as being both a family member and a school classmate. Multiplex edges create overlaps between different social contexts, and as such, are important structures contributing to the overlap part of excess closure. We observe that the share of such multiplex edges in the sampled ego networks is showing a strongly decreasing trend in Figure~\ref{fig:global_properties}F. Second, following \cite{bidart2022analysing} in Figure~\ref{fig:global_properties}G and H, we capture mobility by the average distance of alters (Figure~\ref{fig:global_properties}G), and the share of alters who live in the same municipality as the ego (Figure~\ref{fig:global_properties}H). Both show an increase in the spatial dispersion of ego networks: average geographic distance is increasing, while the share of alters in the same municipality is decreasing over time.

From these averages, it is not possible to determine whether the observed decreasing closure levels -- in terms of both the decreasing average local clustering coefficient and the decreasing average excess closure --  can be attributed to group-level changes such as increasing migration or an aging society, or to individual-level changes in the ego network structure. Moreover, if we look at the change in the population's mean excess closure in various demographic groups comparing 2021 to 2011 in Figure~\ref{fig:demogr_ec_2011_2021}, we see that it decreases almost regardless of age group, education level, migrant generation, gender, or household income quartile. In the next two subsections, we investigate whether this decrease in closure is a group-level or an individual-level phenomenon using a decomposition and an ego network timeline clustering technique.

\subsection{Decreasing closure as group-level or individual change}

Further investigating the overall decline in cohesion demonstrated in the previous subsection, we next ask whether this trend reflects
(i) group-level demographic shifts in the population related to aging or increasing migration or (ii) structural change within individuals' ego networks.
To separate these mechanisms, we use a standard decomposition technique \parencite{kitagawa1955components} of the year-by-year change in average excess closure, $\Delta C^{exc}_t = C^{exc}_t - C^{exc}_{t-1}$, into three components (see Section~\ref{sec:mm} for methodological details):
\begin{itemize}
    \item a term related to population composition changes that captures changes in \emph{group shares} holding group-level closure fixed,
    \item a within-group \emph{individual change} term that captures shifts in the average closure of demographic groups holding their population shares fixed,
    \item and a remaining \emph{interaction} term.
\end{itemize}

\begin{figure}[h!]
    \centering
    \includegraphics[width=\linewidth]{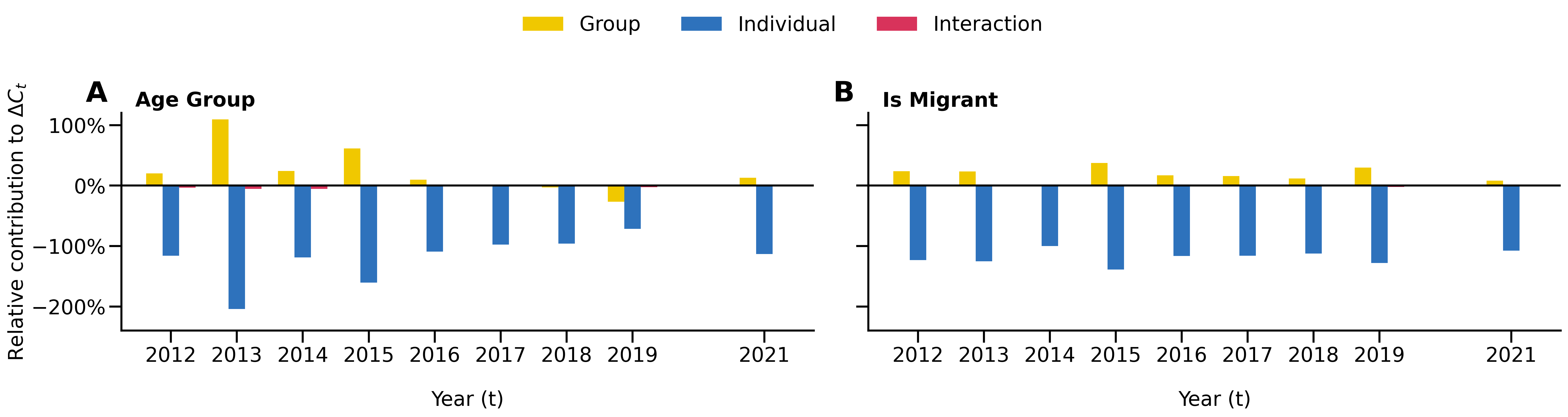}
    \caption{\textbf{Relative contribution to decreasing closure.} Signed relative contributions to the average change in excess closure from $t-1$ to $t$ of the group, individual, and interaction terms (A) by age group, (B) migrant status.}
    \label{fig:pop_vs_ind}
\end{figure}

We report these as signed relative \textit{group}, \textit{individual}, and \textit{interaction} contributions, which add up to the 100\% of total average change in excess closure year by year. Reflecting the two major demographic shifts, Figures~\ref{fig:pop_vs_ind}A and B show the terms after grouping the population by age groups and by migration background. For details , see the \nameref{sec:mm} section. In the year 2020, the average change in excess closure is slightly positive but around 0, therefore, we exclude this year from the timelines.  
Figure~\ref{fig:demogr_comp_2011_2021} presents the detailed shares of the different groups in the population, whereas Figure~\ref{fig:demogr_ec_2011_2021} the groups' average levels of excess closure in 2011 and 2021. 

What we observe from this decomposition is that group-level demographic changes because of aging and increasing share of migrants have positive relative contributions, actually leading to an \textit{increase} in cohesion. On the contrary, individual contributions are almost exclusively negative, outweighing the positive contributions from the group-level shifts. Interaction contributions are negligible. Thus, the decomposition suggests that year-by-year changes are due to ego network structure changes at the individual-level, rather than due to changing group-level demographics. Therefore, as a next step, we turn to the investigation of 12-year timelines of ego networks to further unpack the changing dynamics of ego network structures.

\subsection{Timelines of individual-level degree and closure}

To understand social cohesion change over time, we turn to investigating ego network structure dynamics of individual timelines of normalized degree and normalized excess closure. We apply $k$-means clustering on the timelines that groups individuals whose structural ego network properties follow similar trajectories.  As such, we identify the most common shapes of individual degree and excess closure timelines. For details on the normalization and $k$-means, see \nameref{sec:mm}.

\begin{figure}[h!]
    \centering
    \includegraphics[width=\linewidth]{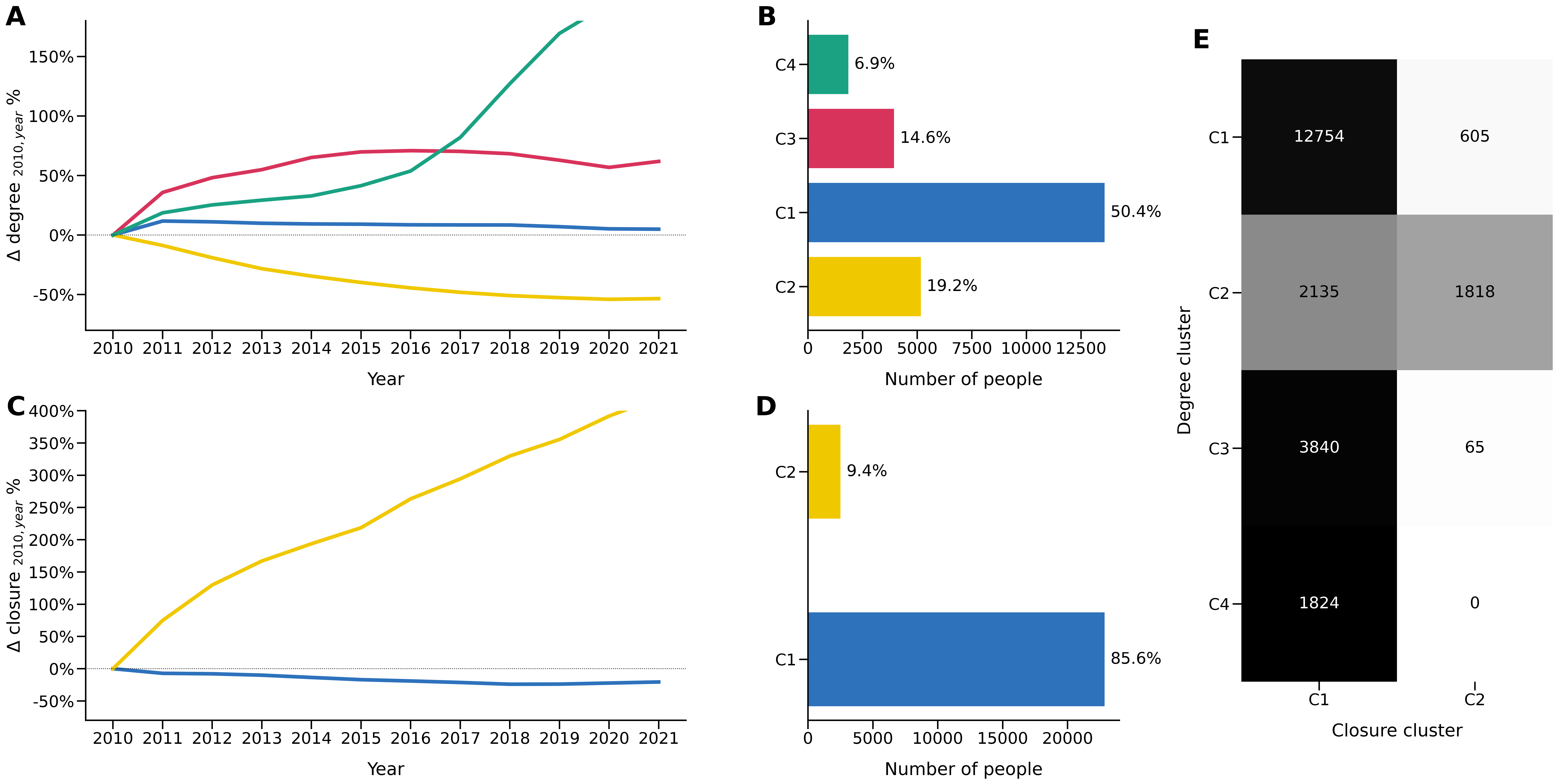}
    \caption{\textbf{Four largest clusters of ego network degree and closure trajectory shapes (indicated by C1--C4 and colors).} (A) Normalized degree ($k_{u,t}^{norm}$) timelines. (B) Share of sample belonging to the normalized degree timeline clusters. (C) Normalized closure ($C_{u,t}^{exc,norm}$) timelines. (D) Share of sample belonging to the normalized closure timeline clusters. (E) Cross-membership between degree (rows) and closure (columns) clusters.}
    \label{fig:kmeans}
\end{figure}

In Figures~\ref{fig:kmeans}A--B, we show all clusters that contain at least 5\% of the individuals, four in total. The most frequent degree pattern (cluster C1) is a stable degree over time, and it applies to roughly half of the sample. The second most common pattern (C2) is a decreasing degree trajectory, observed for about 19\% of individuals. Two further clusters (C3 and C4) capture smaller, substantially less frequent types of increasing degree trajectories.

Panels C and D of Figure~\ref{fig:kmeans} present the excess closure timeline clusters. Focusing on the two largest clusters, we find that the largest cluster of 86\% of the sample shows decreasing excess closure, followed by a smaller group (9.4\%) with increasing excess closure. Previous work has shown that excess closure tends to decrease as degree increases \parencite{bokanyi2023anatomy}. The fact that the share of timelines with decreasing excess closure is larger than the share of timelines with increasing degree suggests that closure also declines in many ego networks where degree is stable.

To confirm this, we combine degree and closure dynamics in Figure~\ref{fig:kmeans}E, which shows how individuals are distributed across the degree (rows) and excess-closure (columns) clusters of A--D. The most numerous group of individuals is that of stable degree who belong to the decreasing excess closure cluster in the top left corner. This indicates that many people keep a similar number of alters over time, while the structure of their ego networks becomes less tightly knit. This reflects that nontrivial structural changes within individual ego networks add up to the decline in closure we observe in the averages. In the next subsection, we turn to finding the potential individual-level drivers of decreasing excess closure in ego networks.

\clearpage

\subsection{Drivers of individual change in excess closure}

To model the drivers of individual change in excess closure  year by year (from year $t-1$ to $t$, where $t\in\left\{2011,\dots,2021\right\}$), we use two-way fixed effect OLS models. In our models, the dependent variable is the percentage change in excess closure of an individual from one year to another ($\Delta C_{u,t}^{exc}$). We first include structural ego network metrics as baseline variables, and then gradually add explanatory variables for multiplexity and geographic mobility in subsequent models. The two-way fixed effect setup separates temporal trends (time fixed effects) and controls for individual characteristics that are constant over time (individual fixed effects) to capture patterns in within-individual temporal changes.  We include change variables from year $t-1$ to $t$ as dependent variables, as well as lagged variables from year $t-1$, and control variables which may change over time. The regression coefficients, $p$-values and the $R^2$ of the models are shown in Table~\ref{tab:reg_results}. For further details on the dependent variable, independent variables, and controls, see \nameref{sec:mm}.

\subsubsection{Network structure as a baseline model}

We begin with Model~1, where we include variables on the ego network structure: the percentage change in total degree ($\Delta k_{u,t}$), change in the number of layers $\Delta l_{u,t}$, and the lagged variables degree ($k_{u,t-1}$), number of layers ($l_{u,t-1}$), excess closure ($C_{u,t-1}^{exc}$), and demographic controls. As expected from the relationship between degree and excess closure \parencite{bokanyi2023anatomy}, percentage change in degree is negatively  and significantly ($p<0.001$) linked to percentage change in excess closure: an 1\% increase in degree change corresponds to -0.5\% decrease in excess closure change, given all other structural factors (change in number of layers, and initial degree, closure, and number of layers) are the same for two individuals. The strongest predictor is the change in the number of network layers. An increase in the number of layers (e.g., because of leaving school or work) is associated with a large and significant ($p<0.001$) decrease (-127.6\% per additional layer) in the closure change. Conversely, one less layer is associated with a significant increase in closure change. As for the initial variable levels, there is small positive relationship (significant at $p<0.001$) with the lagged degree, with each additional initial degree corresponding to a 0.6\% increase in excess closure. The previous level of excess closure and previous level of number of layers both have a positive significant coefficient: individuals starting with very high levels of closure and activity in all layers are more likely to see an increase in the following year.

\begin{landscape}

\begin{table}[!p] 
\centering 
\small 
\begin{tabular}{@{\extracolsep{0pt}}l
                D{.}{.}{-2}
                D{.}{.}{-2}
                D{.}{.}{-2}
                D{.}{.}{-2}
                D{.}{.}{-2}} 
\\[-1.8ex]\hline 
\hline \\[-1.8ex] 
 & \multicolumn{5}{c}{\textit{Dependent variable:}} \\ 
\cline{2-6} 
\\[-1.8ex] 
 & \multicolumn{5}{c}{\% change in excess closure ($t-1 \to t$)} \\ 
\\[-1.8ex] 
 & \multicolumn{1}{c}{\textbf{Model 1}} 
 & \multicolumn{1}{c}{\textbf{Model 2}} 
 & \multicolumn{1}{c}{\textbf{Model 3}} 
 & \multicolumn{1}{c}{\textbf{Model 4}} 
 & \multicolumn{1}{c}{\textbf{Model 5}} \\[0.25ex]
\hline \\[-1.8ex] 

$\Delta$ Degree \% ($t-1 \to t$) 
  & -0.509^{***} & -0.506^{***} & -0.415^{***} & -0.530^{***} & -0.422^{***} \\ 
  & (0.094) & (0.093) & (0.088) & (0.098) & (0.089) \\[0.5ex] 

$\Delta$ Number of Layers ($t-1 \to t$)
  & -1.276^{***} & -1.314^{***} & -1.265^{***} & -1.176^{***} & -1.206^{***} \\ 
  & (0.170) & (0.170) & (0.166) & (0.162) & (0.159) \\[0.5ex]

$\Delta$ Share Multiplex Ties ($t-1 \to t$) 
  &  & 34.433^{***} &  &  & 30.004^{***} \\ 
  &  & (7.364) &  &  & (7.434) \\[0.5ex] 

$\Delta$ Alter \% Same Municipality ($t-1 \to t$) 
  &  &  & 3.146^{***} &  & 3.012^{***} \\ 
  &  &  & (0.783) &  & (0.791) \\[0.5ex] 

Ego Relocation (municipality, $t-1 \to t$) 
  &  &  &  & 0.683^{**} & 0.640^{**} \\ 
  &  &  &  & (0.481) & (0.461) \\[0.5ex] 

Alter Relocation \% (municipality, $t-1 \to t$) 
  &  &  &  & 0.776^{*} & 0.684^{*} \\ 
  &  &  &  & (0.647) & (0.648) \\[0.5ex] 
  
\hline\\[0.25ex]

Degree (t-1) 
  & 0.006^{***} & 0.006^{***} & 0.006^{***} & 0.006^{***} & 0.006^{***} \\ 
  & (0.002) & (0.002) & (0.002) & (0.002) & (0.002) \\[0.5ex] 

Log$_{10}$ Closure (t-1) 
  & 3.766^{***} & 3.799^{***} & 3.895^{***} & 3.789^{***} & 3.941^{***} \\ 
  & (0.730) & (0.756) & (0.832) & (0.743) & (0.862) \\[0.5ex] 

Number of Layers (t-1) 
  & 0.149 & 0.176 & 0.032 & 0.152 & 0.069 \\ 
  & (0.199) & (0.192) & (0.210) & (0.197) & (0.200) \\[0.5ex] 

Share Multiplex Ties (t-1) 
  &  & -23.302^{***} &  &  & -20.867^{***} \\ 
  &  & (8.343) &  &  & (7.969) \\[0.5ex] 

Alter \% Same Municipality (t-1) 
  &  &  & -2.590^{***} &  & -2.461^{***} \\ 
  &  &  & (0.897) &  & (0.851) \\[0.5ex] 

\hline \\[-1.8ex] 
Adjusted R$^{2}$ 
  & \multicolumn{1}{c}{0.038} 
  & \multicolumn{1}{c}{0.039} 
  & \multicolumn{1}{c}{0.042} 
  & \multicolumn{1}{c}{0.039} 
  & \multicolumn{1}{c}{0.043} \\ 
\hline 
\hline \\[-1.8ex] 
\textit{Note:}  
  & \multicolumn{5}{r}{Standard errors in parentheses. $^{*}p<0.05$; $^{**}p<0.01$; $^{***}p<0.001$.} \\ 
\end{tabular} 
\caption{\textbf{Two-way fixed effect panel OLS results} for the percentage change in excess closure between $t-1$ and $t$. Change variables and lagged variables are separated by a horizontal line.} 
\label{tab:reg_results} 
\end{table}
\end{landscape}

\subsubsection{Multiplexity of connections}

From here on, Model~1 serves as a baseline for comparison when investigating the effects of further variables. Model~2 introduces multiplexity in addition to the structural variables. From its coefficients, we see that given the same structure, a 1\% increase in the share of multiplex ties translates to a 34.4\% significant increase in excess closure change, showing that closure is very sensitive to the presence of overlapping social opportunity structures, like neighbors also being school classmates or family members working together. Conversely, if the share of these multiplex ties is decreasing in ego network, it contributes to sharply decreasing closure levels. Initial levels of multiplexity the previous year indicate a regression-to-the-mean effect --- for two individuals with the same ego network structure and the same degree and multiplexity percentage changes, 1\% higher initial multiplexity is associated with a -23.3\% significant ($p<0.001$) percentage decrease in excess closure. Thus, the decreasing average multiplexity we observe in Figure~\ref{fig:global_properties}1F is a strong driver of decreasing closure.

\subsubsection{Geographical dispersion and relocation}

Model~3 includes a change and a lagged variable reflecting the geographic dispersion of alters in addition to the previous variables. We find that a 1\% increase in the share of local alters (alters in the same municipality as the ego) is associated with a significant 3.1\% increase in closure. This confirms that local proximity is a powerful facilitator of triadic closure, and that more geographic dispersion, i.e., less local alters leads to more open ego networks. We also note that including these dispersion variables slightly attenuates the effect of degree change, suggesting that part of the impact of adding/losing ties is explained by where those ties are located. The geographic dispersion of ego networks is related on the one hand to short-term mobility - people living elsewhere than their work colleagues or school classmates. On the other hand, it is influenced by long-term mobility decisions such as relocations, since in this data, relocating individuals may find themselves further away from their family ties. 

We add recent relocations as a dummy variable to both egos and alters in Model~4. Here, we find a 'relocation paradox': \emph{ceteris paribus}, individuals who change their municipality from one year to another have almost 68\% higher closure change (at significance $p<0.01$). The share of relocating alters also matters, as 1\% more alters relocating is associated with a 0.7\% higher closure (with $p<0.1$). It should be noted here that we did not differentiate between alters moving into or out of the municipality of the ego. As such, individuals in general move towards more closed network structures in the year after the relocation, where local connections such as school or work create more network closure. However, relocation also contributes to increasing geographic dispersion in the long run.

Finally, Model~5 includes all variables simultaneously, primarily to test robustness. The coefficients for change in the share of multiplexity and changing geographical dispersion, as well as the relocation variables remain stable and significant. This demonstrates that each of these factors (multiplexity, dispersion, and relocation) plays a distinct and independent role in driving the evolution of individual network closure.

\section{Discussion}

Using twelve annual snapshots of a population-scale, multilayer register network of the Netherlands (2010--2021), we find evidence of decreasing social cohesion: while individuals maintain a remarkably stable number of formal opportunity ties, their ego networks become substantially less cohesive, with average excess closure declining by more than 15\%. Importantly, the decline in closure is not explained by group-level demographic shifts such as aging of the population or immigration. Our decomposition and ego-network timeline analyses indicate that such changes in population composition would, if anything, shift closure upward. Instead, the observed decline arises from individual-level changes in the ego network structures. 

Two individual-level processes account for most of the decline: reduced multiplexity and greater geographic mobility. Multiplex relationships, such as family members who are also co-workers or neighbors, are diminishing, driving closure down. Simultaneously, reflecting changes in mobility patterns, urban structure, and digital infrastructure, the geographic dispersion of social ties is increasing, again contributing to the decline in closure. While online platforms and remote services now allow people to live further away from their schools, workplaces, or family members, these tools may not fully substitute for the multiplexity and local embeddedness that foster strong community bonds.

One particularly striking pattern is the relocation paradox. Moving, typically viewed as disruptive to social networks, is associated with a short-term increase in closure, likely reflecting efforts of individuals to move to the place where they have more local ties. Yet over time, repeated moves and their contribution to increasing geographic dispersion reduce the likelihood of strong, multiplex relationships. As such, mobility can boost local cohesion in the short term for an individual, but cumulatively, it may contribute to more fragmented social networks at the population level.

Substantively, our results are consistent with a broad set of contemporary pressures that can reshape the opportunity structure for social connections. For instance, tight housing markets may force people to move, while suburbanization and longer commuting patterns reduce repeated co-presence in shared social contexts. Additionally, increasing digitalization lowers the costs of sustaining relationships across distance, thereby reducing the prevalence of geographically co-located, overlapping contexts of social life. At the same time, the observed structural shift does not currently map onto attitudinal indicators of cohesion: recent survey evidence for the Netherlands does not show a clear decline in trust or participation \parencite{schmeets2014declining}. A plausible interpretation is that individuals can maintain perceived trust and civic engagement while their networks become less locally clustered. For example, by substituting local overlap with geographically dispersed ties or with digitally mediated interactions that are not fully captured by register-based formal ties. Nevertheless, a more open network structure can already have tangible consequences for informal, place-based support: when alters are less connected to one another and more spatially dispersed, coordinating practical help becomes harder, especially in times of crisis, potentially increasing vulnerability among groups such as elderly or those with limited mobility.

Our analysis is based on formal, register-based ties being family, household, school, work, and next-door neighbor relationships, which presents both a strength and a limitation. On the one hand, these relations capture the institutional backbone of social connectivity: they represent social opportunity structures for repeated co-presence and interaction, are measured consistently for the full population over time, and can be linked to high-quality demographic and geographic information. On the other hand, they provide an incomplete representation of social life, as they do not explicitly list informal friendships or voluntary associations, even if a large share of these stem from previous formal relationships \parencite{vaneijk2010unequal}. Although there is substantial evidence that, at the aggregate level, register-based networks reproduce key connectivity and community patterns observed in an online social network of the same population \parencite{menyhert2025connectivity}, an important open question is how these unobserved informal relational contribute to network closure. Future work that triangulates register-based networks with survey-based ego networks and/or platform-based digital traces would be particularly valuable for assessing how changes in social opportunity structures translate into changes in experienced support, local embeddedness, and perceived cohesion.

Finally, we note the challenge of causal inference. While we identify significant associations between multiplexity, geographical dispersion, relocation, and change in network closure, reverse causality remains possible: individuals with less clustered networks may be more prone to move, or less likely to opt for overlapping social contexts. The individual-level longitudinal analysis helps mitigate some of these concerns, but future quasi-experimental setup or agent-based models could help disentangle cause and effect.

Taken together, our study highlights the importance of understanding the topological evolution of population-scale social networks. As mobility, technology, and economic constraints reshape how people connect, measurable changes in social structure can emerge within a single decade. Our findings on the drivers of declining social cohesion suggest that policy and planning should consider the ongoing fragmentation of locally dense, multiplex communities and the potential social costs of the fragmentation of local, tightly-knit networks.

\clearpage

\section{Data and Methods}
\label{sec:mm}

Below, we discuss the data used in our analysis, as well as the various steps of our methodology for understanding how social cohesion develops over time. 

\subsection{Longitudinal population-scale network}

The longitudinal population-scale social network used in this analysis is the register-based social network of formal ties of all residents of the Netherlands published by Statistics Netherlands \parencite{vanderlaan2022person}. We use five different types of ties: family, household, work, school and next-door neighbor derived yearly between 2010 and 2021 for all residents officially registered in the country on 1st January of the given year. The number of nodes in the yearly networks ranges from 16,577,672 in 2010 to 17,474,700 in 2021, which is also reflected in Figure~\ref{fig:global_properties}A (for detailed numbers for all years, see SI Table~\ref{tab:global_properties}.)

The five different edge types, which can also be viewed as five layers in a multilayer network \parencite{bokanyi2023anatomy}, are based on multiple different country-level registers.
\begin{itemize}
    \item \textbf{Family} edges are derived from the child-parent and partner registers. They include 
    \begin{itemize}
        \item parent--child,
        \item co-parent,
        \item partner,
        \item sibling,
        \item grandparent--grandchild,
        \item aunt/uncle--niece/nephew,
        \item cousin,
        \item stepparent--stepchild,
        \item parent-in-law--child-in-law, and
        \item sister/brother-in-law relationships. 
    \end{itemize}
  
    Parent--child relations are based on the legal definition, and they are only available for parent--child pairs where both people have been recorded in the central register (Gemeentelijke Basisadministratie) after the introduction of digital registers on 1st October 1995. This means a portion of missing family edges for upward and horizontal kin for older residents, see Figure~\ref{fig:share_missing_parents} on the estimated proportion of missing family links for each year. While some family relationships are directed, when all of them are included in the family layer, the layer becomes undirected.
    \item \textbf{Household} edges are based on the address register, there is a household edge between two nodes if they shared address on 1st January the given year. Institutional households (e.g. care homes) are excluded from this analysis.
    \item \textbf{Work} edges are derived from first determining the employer for the main source of income of a person each month in the year leading up to 1st January. Then all co-workers for the same employer are listed for the same month, and the listed co-workers are unioned for the whole year. Employers are only registered at the level of headquarters by the state. Therefore, if the total number of co-workers exceed 100 in total, then the 100 geographically closest colleagues (by address) are sampled to avoid a high number of co-workers for large employers with geographically dispersed locations; this sampling is done independently each year. The sampling introduces directed edges which we consider to be undirected.
    \item \textbf{School} edges are between nodes who visit the same educational institution (primary, secondary, vocational, special or higher education) are in the same year group, and in the case of higher education, follow the same course.
    \item \textbf{Next-door neighbors} are all residents of the 10 closest addresses to the given node's address. If there are multiple households with the same distance, e.g. for large apartment blocks at the same geographical location, 10 households are sampled; this sampling is done independently for each year. Connections to institutional households and random neighborhood connections originally provided in the data as created by Statistics Netherlands are not included in our analysis.
\end{itemize}

In addition to standard tools, we use the \texttt{mlnlib} Python library \parencite{bokanyi2023popnet} for the multilayer network analysis.

\subsection{Node-level attributes}

Alongside the social ties, detailed node-level demographic information is available for individuals each year from Statistics Netherlands microdata. We use the following in our analysis:
\begin{itemize}
    \item \textbf{age} corresponds to the age of individuals in years on 1st January in a given year;
    \item \textbf{gender} is the legal gender as stored in the central register (male or female);
    \item \textbf{highest achieved level of education} of an individual takes the following values: primary, VMBO (secondary applied), HAVO/WVO (secondary scientific), MBO (vocational), WO (higher education);
    \item \textbf{migrant status} is a binary variable indicating whether someone is a first-generation migrant (1), or native/second-generation migrant (0);
    \item \textbf{household income quartile} is calculated for individuals in non-institutional households based on the percentile distribution of the average household income of main earners living at the same address;
    \item \textbf{location} of residential address on 1st January the given year at the neighborhood (buurt) level sourced from the official address register, individuals' coordinates correspond to the centroid of the neighborhood;
    \item \textbf{municipality} derived from neighborhood codes of residential addresses.
    
\end{itemize}

\subsection{Ego network metrics and sampling}
\label{sec:egonet_and_sampling}

We introduce the abstract description of our network dataset following the terminology set out by \textcite{kivela2014multilayer} and also adopted in \cite{bokanyi2023anatomy}. 
We represent the data as an undirected temporal multilayer graph $G_t = (V_t,E_t,L)$, where $V_t$ is the set of $n_t = |V_t|$ nodes in year $t$. The set of undirected edges $E_t$ connects pairs of nodes according to the above described relationships: \[E_t \subseteq \left\{\left(\left\{u,v\right\},\ell\right) : u,v\in V_t, u \neq v, \ell\in L\right\},\] 

There are five layers $L = \{F,H,W,S,N\}$ (corresponding to family (F), household (H), work (W), school (S), and next-door neighbors (N)).

Our multilayer graph is \emph{node-aligned}: nodes are identical across all layers, and there are exclusively intra-layer edges. We can represent the edges with the \emph{reduced adjacency tensor} $A_{uv\ell,t}$, that is a generalization of the adjacency matrix, and encodes edges in the following manner:
\begin{equation}
    A_{uv\ell, t}=\left\{\begin{array}{ll}
    1 & \mbox{if }u\mbox{ and }v\mbox{ are connected in layer }\ell\in L\mbox{ in year }t,\\
    0 & \mbox{otherwise.}
\end{array}\right.
\end{equation}
 
Because the network is undirected, for all $u, v \in V_t$ it holds that $A_{uv\ell,t} = A_{vu\ell,t}$. 
We can also think about our network as an edge-coloured multigraph, that allows multiple types of relationships between two nodes. This representation can be viewed as a "flattened" version of the multilayer representation. 

The \textbf{degree} $k_{u\ell,t}$ of a node $u \in V_t$ in a given layer $\ell\in L$ at time $t$ is given by
\begin{equation}
    k_{u\ell,t} = \sum_{v \in V} A_{uv\ell,t}. 
\end{equation}

The \textbf{total degree} for a node $u$ across all layers at time $t$ is 
\begin{equation}
    k_{u,t} = \sum_{\ell\in L} k_{u\ell,t} = \sum_{\ell\in L} \sum_{v \in V_t} A_{uv\ell,t}.
\end{equation}
Here, we count edges that are present in more layers multiple times. \\
The \textbf{average total degree} is given by\[\left<k_t\right>=\frac{1}{n_t}\sum_{u\in V_t}k_{u,t}.\]

From here on in our analysis, we focus on the topological network properties of the ego networks of a random sample of 50k individuals. The sampling was done from the pool of all people who are present at least once in the yearly networks between 2010 and 2021. Out of the full 50k sample, the number of people in the population each year, and the number of times an individual is in the sample can be found in Table~\ref{tab:sample_size} and Figure~\ref{fig:years_in_sample}.

The \textbf{number of unique neighbors} connected to node $u$ at time $t$ is
\begin{equation}
    k'_{u,t} = \left|\left\{v\in V_t: \sum_{\ell\in L}A_{uv\ell,t} >0 \right\}\right|.
\end{equation}

The \textbf{number of layers} node $u$ is active in at time $t$ equals the number of layers it has at least one  alter in, i.e., 
\begin{equation}
    l_{u,t} = \left|\left\{\ell\in L: \sum_{v\in \mathrm{neighbors}(u)} A_{uv\ell,t} >0 \right\}\right|.
\end{equation}

The \textbf{multiplexity} of an edge between node $u$ and node $v$ at time $t$ is
\begin{equation}
    m_{uv,t} = \sum_{\ell\in L} A_{uv\ell,t}.
\end{equation}

Using multiplexity, we can define the \textbf{share of multiplex edges} around an ego $u$ by calculating the share of $m_{uv,t}$ being larger than 1 out of all unique neighbors:
\begin{equation}
    \mu_{u,t} = \frac{\sum_{v \in \mathrm{neighbors}(u)} \left(m_{uv,t}>1\right)}{k'_{u,t}}.
\end{equation}

The local \textbf{clustering coefficient} $C_{u,t}$ is calculated for each node $u$ in each year from following adjacency matrix $A'_{t}$:
\begin{equation}
    A'_{uv,t}=\left\{\begin{array}{ll}
    1 & \mbox{if }\sum_{l\in L}A_{uv\ell,t} > 0,\\
    0 & \mbox{otherwise.}
\end{array}\right.
\end{equation}

\textbf{Excess closure} ($C_{ut}^{\text{exc}}$) is a measure of clustering developed for multilayer register-based networks in \cite{bokanyi2023anatomy}. It captures closure from inter-layer triangles, but not from intra-layer triangles, which are abundant in social networks based (partly) on underlying bipartite structures. These inter-layer triangles arise either from multiplex edges, where two individuals are connected by a social tie present in multiple layers (such as a family member also being a colleague, or a neighbor also being a classmate), or local bridging where for example, two colleagues are also neighbors of each other. As such, it does not suffer from the problem that the clustering coefficient in register-based networks is inflated because of the presence of intra-layer triangles, and it decreases with increasing degree, which is the expected behaviour from a closure metric observed in online social networks. For more details, we refer the reader to \parencite{bokanyi2023anatomy}.

\subsection{Estimating group-level vs. individual-level contributions to closure change}

To estimate what proportion of the observed decrease in the average closure is because of change in population composition, or because of individual-level changes in closure, we decompose the average year-by-year change as follows \parencite{kitagawa1955components}. For the sake of simplicity, we denote average excess closure in year $t$ by $C_t$ instead of $C^{exc}_t$, where $C_t = \frac{1}{n_t}\sum_{u\in V_t} C_{u,t}$, if $C_{u,t}$ is the closure of individual $u$ in year $t$. The fraction of population in a demographic group $g$ in year $t$ is denoted by $f_{g,t}$, $\sum_g f_{g,t} =1$. Average closure of group $g$ in year $t$ is $C_{g,t} = \frac{1}{n_t\cdot f_g} \sum_{u \in g} C_{u,t}$. Changes from year $t-1$ to $t$ in population fraction is $\Delta f_{g,t} = f_{t}-f_{t-1}$, and $\Delta C_{g,t} = C_{g,t} - C_{g,t-1}$. See fractions for 2011 and 2021 in Figure~\ref{fig:demogr_comp_2011_2021}.

\begin{align*}
C_t &= \sum_g f_{g,t}\cdot C_{g,t}\\
C_{t-1} &= \sum_g f_{g,t-1}\cdot C_{g,t-1}\\
\Delta C_t = C_t - C_{t-1} &= \sum_g f_{g,t}\cdot C_{g,t} - \sum_g f_{g,t-1}\cdot C_{g,t-1} =\\
&= \sum_g \left(f_{g,t-1} + \Delta f_{g,t}\right)\cdot \left(C_{g,t-1}+\Delta C_{g,t}\right)- \sum_g f_{g,t-1}\cdot C_{g,t-1} =\\
&=\underbrace{\sum_g f_{g,t-1}\cdot \Delta C_{g,t}}_{\mbox{within-individual changes}} + \underbrace{\sum_g \Delta f_{g,t}\cdot C_{g,t-1}}_{\mbox{population composition change}} + \underbrace{\sum_g \Delta f_{g,t}\cdot \Delta C_{g,t}}_{\mbox{interaction term}}
\end{align*}

The relative share of the three terms are defined as:
\begin{align}
    S^{\mathrm{group}}_t &= \frac{\sum_g \Delta f_{g,t}\cdot C_{g,t-1}}{|\Delta C_t|},\\
    S^{\mathrm{individual}}_t &= \frac{\sum_g f_{g,t-1}\cdot \Delta C_{g,t}}{|\Delta C_t|},\\
    S^{\mathrm{interaction}}_t &= \frac{\sum_g \Delta f_{g,t}\cdot \Delta C_{g,t}}{|\Delta C_t|}.
\end{align}

Note that while $S^{\mathrm{group}}_t + S^{\mathrm{individual}}_t + S^{\mathrm{interaction}}_t=1$, the terms can both be negative and positive, and as such, can take any real values. We calculate $S$ values for all years, with the grouping corresponding to age groups and migrant status. See Figure~\ref{fig:demogr_comp_2011_2021} and Figure~\ref{fig:demogr_ec_2011_2021} for group shares and group excess closure averages in 2011 and 2021.

\subsection{Timeline clustering with $k$-means}

To cluster the degree and closure timelines of ego networks, we first normalize each degree and closure value to compare percentage change with respect to the initial year 2010. Thus, \textbf{normalized degree} is
\[k^{norm}_{u,t} = \frac{k_{u,t}-k_{u,2010}}{k_{u,2010}},\]
and the \textbf{normalized closure} is
\[C^{exc,norm}_{u,t} = \frac{C^{exc}_{u,t}-C^{exc}_{u,2010}}{C^{exc}_{u,2010}}.\]

Based on the normalized measures, the degree and closure timelines can be viewed as vectors of length 12, from years 2010 to 2021.

We cluster these timeline shapes using the well-known $k$-means algorithm as implemented in the \texttt{scikit-learn} package~\parencite{pedregosa2011scikitlearn}, for which we select the optimal number of clusters using the elbow method. The optimal number of clusters for the degree timelines is 9, and for the closure timelines 10. We then take the vector corresponding to the cluster centers, and represent the clusters by this ``average'' timeline corresponding to the cluster center. To comply with privacy regulations of the data provider, we only present cluster results where the cluster contains at least 10 people.

\subsection{Regression model and variables}

In our regression models, the dependent variable is yearly within-individual change of excess closure $\Delta C_{u,t} = C_{u,t}-C_{u,t-1}$ for node $u$ in year $t$. We use two groups of explanatory variables and a number of demographic control variables:
\begin{itemize}
    \item Network structure variables
    \begin{itemize}
        \item degree in the previous year $k_{u,t-1}$,
        \item change in degree as compared to the previous year $\Delta k_{u,t} = k_{u,t} - k_{u,t-1}$,
        \item excess closure in the previous year $C^{exc}_{u,t-1}$,
        \item change in excess closure as compared to the previous year $\Delta C^{exc}_{u,t} = C^{exc}_{u,t} - C^{exc}_{u,t-1}$,
        \item number of network layers in the previous year $l_{u,t-1}$
        \item change in the number of network layers as compared to the previous year $\Delta l_{u,t} = l_{u,t} - l_{u,t-1}$,
        \item share of multiplex ties in the previous year $\mu_{u,t}$,
        \item change in the share of multiplex ties as compared to the previous year $\Delta \mu_{u,t} = \mu_{u,t} - \mu_{u,t-1}$.
    \end{itemize}
    \item Spatial variables
    \begin{itemize}
        \item a dummy variable $R_{u,t}$ indicating whether the ego relocated into a new municipality as compared to the previous year ($R_{u,t}=1$ if the resident municipality of ego in year $t$ is different from the resident municipality of ego in year $t-1$, and 0 otherwise),
        \item the share of alters who relocated into a new municipality as compared to the previous year $\rho_{u,t} = \sum_{v\in \mathrm{neighbors}_t(u)} R_{u,t} / k'_{u,t}$,
        \item the share of local alters the previous year
        \begin{equation}
            \Lambda_{u,t-1} = \frac{\mbox{number of alters in same municipality in year }t-1}{k'_{u,t-1}}.
        \end{equation}
    \end{itemize}
    \item Control variables
    \begin{itemize}
        \item age group at $t-1$ calculated from age as a categorical variable (0-11, 12-17, 18-25, 26-35, 36-45, 46-55, 56-65, 66-75, 76-85, 86-95, 96+), 
        \item household income quartile at $t-1$ as a categorical variable (Q1--Q4), and
        \item highest achieved level of education at $t-1$ as a categorical variable (1 -- primary, 2 -- secondary, 3 -- tertiary).
    \end{itemize}
    
\end{itemize}

We model within-individual changes using two-way fixed effect panel OLS regressions as implemented in~\cite{kevinsheppard2025bashtage}. Entity fixed effects are added for all datapoints belonging to the same node/individual $u$ across the years, and for the time variable $t$.

\printbibliography

@article{asikainen2020cumulative,
  author  = {Asikainen, Aili and Iñiguez, Gerardo and Ureña-Carrión, Javier and Kaski, Kimmo and Kivelä, Mikko},
  title   = {Cumulative effects of triadic closure and homophily in social networks},
  journal = {Science Advances},
  year    = {2020},
  volume  = {6},
  number  = {19},
  pages   = {eaax7310},
  doi     = {10.1126/sciadv.aax7310},
}

@article{bianconi2014triadic,
  author  = {Bianconi, Ginestra and Darst, Richard K. and Iacovacci, Jacopo and Fortunato, Santo},
  title   = {Triadic closure as a basic generating mechanism of communities in complex networks},
  journal = {Physical Review E},
  year    = {2014},
  volume  = {90},
  number  = {4},
  pages   = {042806},
  doi     = {10.1103/physreve.90.042806},
  url     = {https://link.aps.org/doi/10.1103/PhysRevE.90.042806},
  urldate = {2022-08-31},
  month   = oct,
}

@article{bidart2022analysing,
  author  = {Bidart, Claire and Maisonobe, Marion and Viry, Gil},
  title   = {Analysing {Personal} {Networks} in {Geographical} {Space} {Beyond} the {Question} of {Distance}},
  journal = {Social Inclusion},
  year    = {2022},
  volume  = {10},
  number  = {3},
  pages   = {233--247},
  doi     = {10.17645/si.v10i3.5381},
  url     = {https://shs.hal.science/halshs-03704398},
  urldate = {2024-07-08},
  month   = jun,
}

@article{bokanyi2023anatomy,
  author  = {Bokányi, Eszter and Heemskerk, E. M. and Takes, Frank W.},
  title   = {The anatomy of a population-scale social network},
  journal = {Scientific Reports},
  year    = {2023},
  volume  = {13},
  number  = {1},
  pages   = {9209},
  doi     = {10.1038/s41598-023-36324-9},
  url     = {https://www.nature.com/articles/s41598-023-36324-9},
  urldate = {2024-06-21},
  month   = jun,
}

@misc{bokanyi2023popnet,
  author    = {Bokányi, Eszter and Kazmina, Yuliia and de Jong, Rachel},
  title     = {POPNET: MultiLayerNetwork Python Class},
  publisher = {Zenodo},
  year      = {2023},
  doi       = {10.5281/zenodo.10838866},
  url       = {https://zenodo.org/records/10838866},
  urldate   = {2025-06-24},
  note      = {Version v2.0},
}

@incollection{bourdieu1986forms,
  author    = {Bourdieu, Pierre},
  title     = {The forms of capital},
  booktitle = {Handbook of Theory and Research for the Sociology of Education},
  editor    = {Richardson, John G.},
  year      = {1986},
  publisher = {Greenwood Press},
  address   = {New York},
  pages     = {241--258},
}

@article{calabrese2011interplay,
  author  = {Calabrese, Francesco and Smoreda, Zbigniew and Blondel, Vincent D and Ratti, Carlo},
  title   = {Interplay between telecommunications and face-to-face interactions: {A} study using mobile phone data},
  journal = {PLoS ONE},
  year    = {2011},
  volume  = {6},
  number  = {7},
  pages   = {e20814},
  doi     = {10/c67pnf},
  note    = {arXiv:1101.4505},
  eprint  = {1101.4505},
  archivePrefix = {arXiv},
  url     = {https://doi.org/10/c67pnf},
}

@techreport{candogan2025network,
  author      = {Candogan, Ozan and K{\"o}nig, Michael D. and Marray, Kieran and Takes, Frank W.},
  title       = {Network Rewiring and Spatial Targeting: Optimal Disease Mitigation in Multilayer Social Networks},
  institution = {Becker Friedman Institute for Economics, University of Chicago},
  type        = {Working Paper},
  number      = {2025-14},
  year        = {2025},
  doi         = {10.2139/ssrn.5106505},
  url         = {https://bfi.uchicago.edu/wp-content/uploads/2025/01/BFI_WP_2025-14.pdf},
}

@article{coleman1988social,
  author  = {Coleman, James S},
  title   = {Social capital in the creation of human capital},
  journal = {American Journal of Sociology},
  year    = {1988},
  volume  = {94},
  number  = {1},
  pages   = {95--120},
  doi     = {10.1086/228943},
}

@article{cremers2025unveiling,
  author  = {Cremers, Jolien and Kohler, Benjamin and Maier, Benjamin Frank and Eriksen, Stine Nymann and Einsiedler, Johanna and Christensen, Frederik Kølby and Lehmann, Sune and Lassen, David Dreyer and Mortensen, Laust Hvas and Bjerre-Nielsen, Andreas},
  title   = {Unveiling the social fabric through a temporal, nation-scale social network and its characteristics},
  journal = {Scientific Reports},
  year    = {2025},
  volume  = {15},
  number  = {1},
  pages   = {18383},
  doi     = {10.1038/s41598-025-98072-2},
  url     = {https://www.nature.com/articles/s41598-025-98072-2},
  urldate = {2025-07-08},
  month   = may,
}

@article{dulmen2022places,
  author  = {{van Dülmen}, Christoph and Klärner, Andreas},
  title   = {Places {That} {Bond} and {Bind}: {On} the {Interplay} of {Space}, {Places}, and {Social} {Networks}},
  journal = {Social Inclusion},
  year    = {2022},
  volume  = {10},
  number  = {3},
  pages   = {248--261},
  doi     = {10.17645/si.v10i3.5309},
  url     = {https://www.cogitatiopress.com/socialinclusion/article/view/5309},
  urldate = {2022-11-25},
  month   = sep,
}

@book{durkheim1893division,
  author    = {Durkheim, {\'E}mile},
  title     = {De la division du travail social: {\'e}tude sur l'organisation des soci{\'e}t{\'e}s sup{\'e}rieures},
  year      = {1893},
  publisher = {F{\'e}lix Alcan},
  address   = {Paris},
  url       = {https://archive.org/details/deladivisiondutr00durkuoft},
}

@book{durkheim2005suicide,
  author    = {Durkheim, {\'E}mile},
  title     = {Suicide: A study in sociology},
  year      = {2005},
  publisher = {Routledge},
  address   = {London},
}

@article{eagle2009inferring,
  author  = {Eagle, Nathan and Pentland, Alex Sandy and Lazer, David},
  title   = {Inferring friendship network structure by using mobile phone data.},
  journal = {Proceedings of the National Academy of Sciences of the United States of America},
  year    = {2009},
  volume  = {106},
  number  = {36},
  pages   = {15274--15278},
  doi     = {10/c85f22},
  url     = {http://www.pubmedcentral.nih.gov/articlerender.fcgi?artid=2741241&tool=pmcentrez&rendertype=abstract},
  note    = {ISBN: 1091-6490 (Electronic){\textbackslash}r0027-8424 (Linking)},
}

@article{feld1981focused,
  author  = {Feld, Scott L.},
  title   = {The {Focused} {Organization} of {Social} {Ties}},
  journal = {American Journal of Sociology},
  year    = {1981},
  volume  = {86},
  number  = {5},
  pages   = {1015--1035},
  doi     = {10.1086/227352},
  url     = {https://www.journals.uchicago.edu/doi/10.1086/227352},
  urldate = {2025-07-18},
  month   = mar,
}

@book{festinger1950social,
  author    = {Festinger, Leon and Schachter, Stanley and Back, Kurt},
  title     = {Social pressures in informal groups; a study of human factors in housing},
  year      = {1950},
  publisher = {Harper},
  address   = {Oxford, England}
}

@misc{fischer2018uc,
  author    = {Fischer, Claude S.},
  title     = {{UC} {Berkeley} {Social} {Networks} {Study} ({UCNets}), {San} {Francisco} {Bay} {Area}, 2015-2018: {Version} 2},
  publisher = {ICPSR - Interuniversity Consortium for Political and Social Research},
  year      = {2018},
  doi       = {10.3886/icpsr36975.v2},
  url       = {https://www.icpsr.umich.edu/web/NACDA/studies/36975/versions/V2},
  urldate   = {2026-01-19},
}

@article{garcia-bernardo2024assessing,
  author  = {Garcia-Bernardo, Javier and Hedde-von Westernhagen, Christine and Emery, Tom and {van Hoek}, Albert Jan},
  title   = {Assessing {COVID}-19 transmission through school and family networks using population-level registry data from the {Netherlands}},
  journal = {Scientific Reports},
  year    = {2024},
  volume  = {14},
  number  = {1},
  pages   = {31248},
  doi     = {10.1038/s41598-024-82646-7},
  url     = {https://www.nature.com/articles/s41598-024-82646-7},
  urldate = {2025-07-08},
  month   = dec,
}

@article{garcia2025prediction,
  author  = {Garcia-Bernardo, Javier and Jaspers, Eva and Machado, Weverthon and Plach, Samuel and {van Leeuwen}, Erik Jan},
  title   = {Prediction Gaps as Pathways to Explanation: Rethinking Educational Outcomes through Differences in Model Performance},
  journal = {arXiv},
  year    = {2025},
  note    = {arXiv:2506.22993},
  doi     = {10.48550/arXiv.2506.22993},
  url     = {https://arxiv.org/abs/2506.22993},
  eprint  = {2506.22993},
  archivePrefix = {arXiv},
}

@article{hedde2024predicting,
  author  = {Hedde-von Westernhagen, Christine and Bagheri, Ayoub and Garcia-Bernardo, Javier},
  title   = {Predicting COVID-19 infections using multi-layer centrality measures in population-scale networks},
  journal = {Applied Network Science},
  year    = {2024},
  volume  = {9},
  number  = {1},
  pages   = {27},
  doi     = {10.1007/s41109-024-00632-4},
}

@article{holt-lunstad2015loneliness,
  author  = {Holt-Lunstad, Julianne and Smith, Timothy B. and Baker, Mark and Harris, Tyler and Stephenson, David},
  title   = {Loneliness and social isolation as risk factors for mortality: a meta-analytic review},
  journal = {Perspectives on Psychological Science: A Journal of the Association for Psychological Science},
  year    = {2015},
  volume  = {10},
  number  = {2},
  pages   = {227--237},
  doi     = {10.1177/1745691614568352},
  month   = mar,
}

@article{kazmina2024contact,
  author  = {Kazmina, Yuliia and Heemskerk, E. M. and Bok{\'a}nyi, Eszter and Takes, Frank W},
  title   = {From Contact to Threat: A Social Network Perspective on Perceptions of Immigration},
  journal = {arXiv},
  year    = {2024},
  note    = {arXiv:2407.06820},
  doi     = {10.48550/arXiv.2407.06820},
  url     = {https://arxiv.org/abs/2407.06820},
  eprint  = {2407.06820},
  archivePrefix = {arXiv},
}

@article{kazmina2024socioeconomic,
  author  = {Kazmina, Yuliia and Heemskerk, E. M. and Bokányi, Eszter and Takes, Frank W.},
  title   = {Socio-economic segregation in a population-scale social network},
  journal = {Social Networks},
  year    = {2024},
  volume  = {78},
  pages   = {279--291},
  doi     = {10.1016/j.socnet.2024.02.005},
  url     = {https://linkinghub.elsevier.com/retrieve/pii/S0378873324000157},
  urldate = {2024-06-25},
  month   = jul,
}

@article{kazmina2025can,
  author  = {Kazmina, Yuliia and Heemskerk, Eelke M and {van der Kooij}, Emilia and Bokányi, Eszter and Takes, Frank W},
  title   = {Can social capital remedy structural inequality? {Economic} mobility in a longitudinal population-scale social network},
  journal = {arXiv},
  year    = {2025},
  note    = {arXiv:2508.05275},
  doi     = {10.48550/arXiv.2508.05275},
  url     = {https://arxiv.org/abs/2508.05275},
  eprint  = {2508.05275},
  archivePrefix = {arXiv},
}

@misc{kevinsheppard2025bashtage,
  author    = {Sheppard, Kevin and {{{{Snyk bot}}}} and Ro, Joon and Lewis, Brian and Clauss, Christian and Guangyi and Jeff and Yu, Jerry Qinghui and Jiageng and Wilson, Kevin and Migrator, LGTM and Thrasibule and WilliamRoyNelson and RENE-CORAIL, Xavier and vikjam},
  title     = {bashtage/linearmodels: Release 7.0},
  publisher = {Zenodo},
  year      = {2025},
  doi       = {10.5281/zenodo.17407190},
  url       = {https://zenodo.org/doi/10.5281/zenodo.17407190},
  urldate   = {2026-01-19},
  note      = {Version v7.0 (published 2025-10-21)},
  month     = oct,
}

@article{khemakongkanonth2025empirical,
  author  = {Khemakongkanonth, Chate},
  title   = {An empirical analysis on relationships between over-the-top applications for communication and traditional mobile voice services},
  journal = {Telecommunications Policy},
  year    = {2025},
  volume  = {49},
  number  = {3},
  pages   = {102916},
  doi     = {10.1016/j.telpol.2025.102916},
  url     = {https://linkinghub.elsevier.com/retrieve/pii/S0308596125000138},
  urldate = {2026-01-19},
  month   = apr,
}

@article{kitagawa1955components,
  author  = {Kitagawa, Evelyn M.},
  title   = {Components of a Difference Between Two Rates},
  journal = {Journal of the American Statistical Association},
  year    = {1955},
  volume  = {50},
  number  = {272},
  pages   = {1168--1194},
  doi     = {10.1080/01621459.1955.10501299},
}

@article{kivela2014multilayer,
  author  = {Kivela, M. and Arenas, A. and Barthelemy, M. and Gleeson, J. P. and Moreno, Y. and Porter, M. A.},
  title   = {Multilayer networks},
  journal = {Journal of Complex Networks},
  year    = {2014},
  volume  = {2},
  number  = {3},
  pages   = {203--271},
  doi     = {10.1093/comnet/cnu016},
  url     = {https://academic.oup.com/comnet/article-lookup/doi/10.1093/comnet/cnu016},
  urldate = {2022-04-06},
  month   = sep,
}

@book{larsen2013rise,
  author    = {Larsen, Christian Albrekt},
  title     = {The {Rise} and {Fall} of {Social} {Cohesion}: {The} {Construction} and {De}-construction of {Social} {Trust} in the {US}, {UK}, {Sweden} and {Denmark}},
  year      = {2013},
  publisher = {Oxford University Press},
  address   = {Oxford},
  isbn      = {978-0-19-968184-6},
  month     = jun,
}

@article{menyhert2025connectivity,
  author  = {Menyhért, Márton and Bokányi, Eszter and Corten, Rense and Heemskerk, Eelke M. and Kazmina, Yuliia and Takes, Frank W.},
  title   = {Connectivity and community structure of online and register-based social networks},
  journal = {EPJ Data Science},
  year    = {2025},
  volume  = {14},
  number  = {1},
  pages   = {8},
  doi     = {10.1140/epjds/s13688-025-00522-4},
  url     = {https://epjdatascience.springeropen.com/articles/10.1140/epjds/s13688-025-00522-4},
  urldate = {2025-06-18},
  month   = jan,
}

@book{newman2010networks,
  author    = {Newman, M. E. J.},
  title     = {Networks: an introduction},
  year      = {2010},
  publisher = {Oxford University Press},
  address   = {Oxford},
  isbn      = {978-0-19-159417-5},
  doi       = {10.1093/acprof:oso/9780199206650.001.0001},
}

@article{onnela2007structure,
  author  = {Onnela, J.-P. and Saramäki, J. and Hyvönen, J. and Szabó, G. and Lazer, D. and Kaski, K. and Kertész, J. and Barabási, A.-L.},
  title   = {Structure and tie strengths in mobile communication networks},
  journal = {Proceedings of the National Academy of Sciences},
  year    = {2007},
  volume  = {104},
  number  = {18},
  pages   = {7332--7336},
  doi     = {10.1073/pnas.0610245104},
  url     = {https://www.pnas.org/doi/full/10.1073/pnas.0610245104},
  urldate = {2024-07-29},
  month   = may,
}

@article{panayiotou2025anatomy,
	title = {Anatomy of a {Swedish} population-scale network},
	volume = {15},
	issn = {2045-2322},
	url = {https://www.nature.com/articles/s41598-025-15966-x},
	doi = {10.1038/s41598-025-15966-x},
	number = {1},
	urldate = {2026-01-30},
	journal = {Scientific Reports},
	author = {Panayiotou, Georgios and Wohlert, Inga K. and Bask, Miia and Bask, Mikael and Magnani, Matteo and Mäkinen, Ilkka Henrik},
	month = aug,
	year = {2025},
	pages = {30300},
}

@article{park2018strength,
  author  = {Park, Patrick S. and Blumenstock, Joshua E. and Macy, Michael W.},
  title   = {The strength of long-range ties in population-scale social networks},
  journal = {Science},
  year    = {2018},
  volume  = {362},
  number  = {6421},
  pages   = {1410--1413},
  doi     = {10/gfrksw},
  urldate = {2021-09-16},
  month   = dec,
}

@article{pedregosa2011scikitlearn,
  author  = {Pedregosa, Fabian and Varoquaux, Ga{\"e}l and Gramfort, Alexandre and Michel, Vincent and Thirion, Bertrand and Grisel, Olivier and Blondel, Mathieu and Prettenhofer, Peter and Weiss, Ron and Dubourg, Vincent and VanderPlas, Jake and Passos, Alexandre and Cournapeau, David and Brucher, Matthieu and Perrot, Matthieu and Duchesnay, {\'E}douard},
  title   = {Scikit-learn: Machine Learning in Python},
  journal = {Journal of Machine Learning Research},
  year    = {2011},
  volume  = {12},
  pages   = {2825--2830},
  url     = {https://jmlr.org/papers/v12/pedregosa11a.html},
}

@article{peixoto2022disentangling,
  author  = {Peixoto, Tiago P.},
  title   = {Disentangling {Homophily}, {Community} {Structure}, and {Triadic} {Closure} in {Networks}},
  journal = {Physical Review X},
  year    = {2022},
  volume  = {12},
  number  = {1},
  pages   = {011004},
  doi     = {10.1103/physrevx.12.011004},
  url     = {https://link.aps.org/doi/10.1103/PhysRevX.12.011004},
  urldate = {2025-07-18},
  month   = jan,
}

@article{poortinga2007perceptions,
  author  = {Poortinga, Wouter and Dunstan, Frank D and Fone, David L},
  title   = {Perceptions of the neighbourhood environment and self rated health: a multilevel analysis of the Caerphilly Health and Social Needs Study},
  journal = {BMC Public Health},
  year    = {2007},
  volume  = {7},
  number  = {1},
  pages   = {285},
  doi     = {10.1186/1471-2458-7-285},
}

@incollection{putnam2000bowling,
  author    = {Putnam, Robert D.},
  title     = {Bowling {Alone}: {America}’s {Declining} {Social} {Capital}},
  booktitle = {Culture and {Politics}},
  editor    = {L., Crothers and C., Lockhart},
  year      = {2000},
  publisher = {Palgrave Macmillan US},
  address   = {New York},
  pages     = {223--234},
  doi       = {10.1007/978-1-349-62965-7_12},
  url       = {http://link.springer.com/10.1007/978-1-349-62965-7_12},
}

@article{sampson1997neighborhoods,
  author  = {Sampson, Robert J and Raudenbush, Stephen W and Earls, Felton},
  title   = {Neighborhoods and violent crime: A multilevel study of collective efficacy},
  journal = {Science},
  year    = {1997},
  volume  = {277},
  number  = {5328},
  pages   = {918--924},
  doi     = {10.1126/science.277.5328.918},
}

@article{saramaki2014persistence,
  author  = {Saramäki, Jari and Leicht, E. A. and López, Eduardo and Roberts, Sam G. B. and Reed-Tsochas, Felix and Dunbar, R. I. M.},
  title   = {Persistence of social signatures in human communication},
  journal = {Proceedings of the National Academy of Sciences},
  year    = {2014},
  volume  = {111},
  number  = {3},
  pages   = {942--947},
  doi     = {10.1073/pnas.1308540110},
  url     = {https://pnas.org/doi/full/10.1073/pnas.1308540110},
  urldate = {2025-07-08},
  month   = jan,
}

@article{schmeets2014declining,
  author  = {Schmeets, Hans and te Riele, Saskia},
  title   = {Declining {Social} {Cohesion} in {The} {Netherlands}?},
  journal = {Social Indicators Research},
  year    = {2014},
  volume  = {115},
  number  = {2},
  pages   = {791--812},
  doi     = {10.1007/s11205-013-0234-x},
  url     = {https://doi.org/10.1007/s11205-013-0234-x},
  urldate = {2025-05-28},
  month   = jan,
}

@misc{share-eric2024survey-2,
  author    = {SHARE-ERIC},
  title     = {Survey of {Health}, {Ageing} and {Retirement} in {Europe} ({SHARE}) {Wave} 4},
  publisher = {SHARE-ERIC},
  year      = {2024},
  doi       = {10.6103/share.w4.900},
  url       = {https://share-eric.eu/data/data-documentation/waves-overview/wave-4},
  urldate   = {2026-01-19},
}

@book{shaw1942juvenile,
  author    = {Shaw, Clifford R. and McKay, Henry D.},
  title     = {Juvenile Delinquency and Urban Areas: A Study of Rates of Delinquents in Relation to Differential Characteristics of Local Communities in American Cities},
  year      = {1942},
  publisher = {University of Chicago Press},
  address   = {Chicago, IL},
  series    = {Behavior Research Fund Monographs},
}

@book{simmel1964conflict,
  author    = {Simmel, Georg},
  title     = {Conflict and the web of group-affiliations},
  year      = {1964},
  publisher = {Free Press},
  address   = {New York},
}

@article{subramanian2002social,
  author  = {Subramanian, Subu V and Kim, Daniel J and Kawachi, Ichiro},
  title   = {Social trust and self-rated health in US communities: a multilevel analysis},
  journal = {Journal of Urban Health},
  year    = {2002},
  volume  = {79},
  number  = {Suppl 1},
  pages   = {S21--S34},
  doi     = {10.1093/jurban/79.4.401},
}

@book{toennies1887gemeinschaft,
  author    = {Tönnies, Ferdinand},
  title     = {Gemeinschaft und Gesellschaft},
  year      = {1887},
  publisher = {Fues's Verlag},
  address   = {Leipzig},
}

@misc{vanderlaan2022person,
  author  = {{van der Laan}, Jan},
  title   = {A {Person} {Network} of the {Netherlands}},
  address = {CBS Discussion Papers},
  year    = {2022},
  url     = {https://www.cbs.nl/-/media/_pdf/2022/20/person_network_netherlands.pdf},
  month   = may,
}

@book{vaneijk2010unequal,
  author    = {{van Eijk}, G.},
  title     = {Unequal networks: spatial segregation, relationships and inequality in the city},
  year      = {2010},
  publisher = {Delft Univ Press},
  address   = {Delft},
  series    = {Sustainable urban areas},
  number    = {32},
  isbn      = {978-1-60750-555-6},
  note      = {OCLC: ocn645493407},
}

@article{vasquesfilho2020transitivity,
  author  = {Vasques Filho, Demival and O'Neale, Dion R. J.},
  title   = {Transitivity and degree assortativity explained: {The} bipartite structure of social networks},
  journal = {Physical Review E},
  year    = {2020},
  volume  = {101},
  number  = {5},
  pages   = {052305},
  doi     = {10.1103/physreve.101.052305},
  url     = {https://link.aps.org/doi/10.1103/PhysRevE.101.052305},
  urldate = {2025-07-18},
  month   = may,
}

@article{viry2012residential,
  author  = {Viry, Gil},
  title   = {Residential mobility and the spatial dispersion of personal networks: {Effects} on social support},
  journal = {Social Networks},
  year    = {2012},
  volume  = {34},
  number  = {1},
  pages   = {59--72},
  doi     = {10.1016/j.socnet.2011.07.003},
  url     = {https://www.sciencedirect.com/science/article/pii/S0378873311000475},
  urldate = {2024-07-24},
  month   = jan,
}

@article{volker2020social,
  author  = {Völker, Beate},
  title   = {Social capital across the life course: {Accumulation}, diminution, or segregation?},
  journal = {Network Science},
  year    = {2020},
  volume  = {8},
  number  = {3},
  pages   = {313--332},
  doi     = {10.1017/nws.2020.26},
  url     = {https://www.cambridge.org/core/journals/network-science/article/social-capital-across-the-life-course-accumulation-diminution-or-segregation/C609093A26323BEEFDFD23302239C0AE},
  urldate = {2026-01-19},
  month   = sep,
}

@article{wellman1979community,
  author  = {Wellman, Barry},
  title   = {The community question: The intimate networks of East Yorkers},
  journal = {American Journal of Sociology},
  year    = {1979},
  volume  = {84},
  number  = {5},
  pages   = {1201--1231},
  doi     = {10.1086/226906},
}

@book{wellman1999networks,
  editor    = {Wellman, Barry},
  title     = {Networks in the global village: life in contemporary communities},
  year      = {1999},
  publisher = {Westview Press},
  address   = {Boulder, Colorado},
  isbn      = {978-0-8133-6821-4},
}

@article{wellmann2019are,
  author  = {Wellmann, Nicolas},
  title   = {Are {OTT} messaging and mobile telecommunication an interrelated market? {An} empirical analysis},
  journal = {Telecommunications Policy},
  year    = {2019},
  volume  = {43},
  number  = {9},
  pages   = {101831},
  doi     = {10.1016/j.telpol.2019.101831},
  url     = {https://www.sciencedirect.com/science/article/pii/S0308596117303804},
  urldate = {2026-01-19},
  month   = oct,
}

\clearpage

\section*{Acknowledgements}

Eszter Bokányi would like to thank Javier Garcia-Bernardo and Rense Corten for their inspiring suggestions and comments.

\subsection*{Funding information}

The authors acknowledge that they received funding in support for this research from the Dutch Research Council (NWO) through the SSHOC-NL project. 

\subsection*{Data availability statement}

All data needed to evaluate the conclusions in the paper as well as access procedures and further information on the dataset are deposited in the secure storage of the ODISSEI portal\footnote{\href{https://odissei-data.nl/facility/odissei-portal/}{https://odissei-data.nl/facility/odissei-portal/}} in the following repository: \href{https://doi.org/10.34894/8575OP}{https://doi.org/10.34894/8575OP}. Access can be requested after obtaining authorization to use the Statistics Netherlands (CBS) Remote Access (RA) Microdata environment\footnote{\href{https://www.cbs.nl/en-gb/our-services/ customised-services-microdata}{https://www.cbs.nl/en-gb/our-services/customised-services-microdata}}.

\subsection*{Author contributions}

EB, FWT, and EMH conceived the study. EB performed the data work, developed the code and the methodology, led the analyses, and created the figure. YK contributed to the methods and analyses. All authors contributed to writing, reviewed the manuscript, and approved the final version.

\subsection*{Competing interests}

The authors declare that they have no competing interests.

\clearpage

\section{Supplementary Information}

\setcounter{figure}{0}
\setcounter{table}{0}

\renewcommand{\thefigure}{SI\arabic{figure}}
\renewcommand{\thetable}{SI\arabic{table}}

\begin{figure}[h!]
    \centering
    \includegraphics[width=\linewidth]{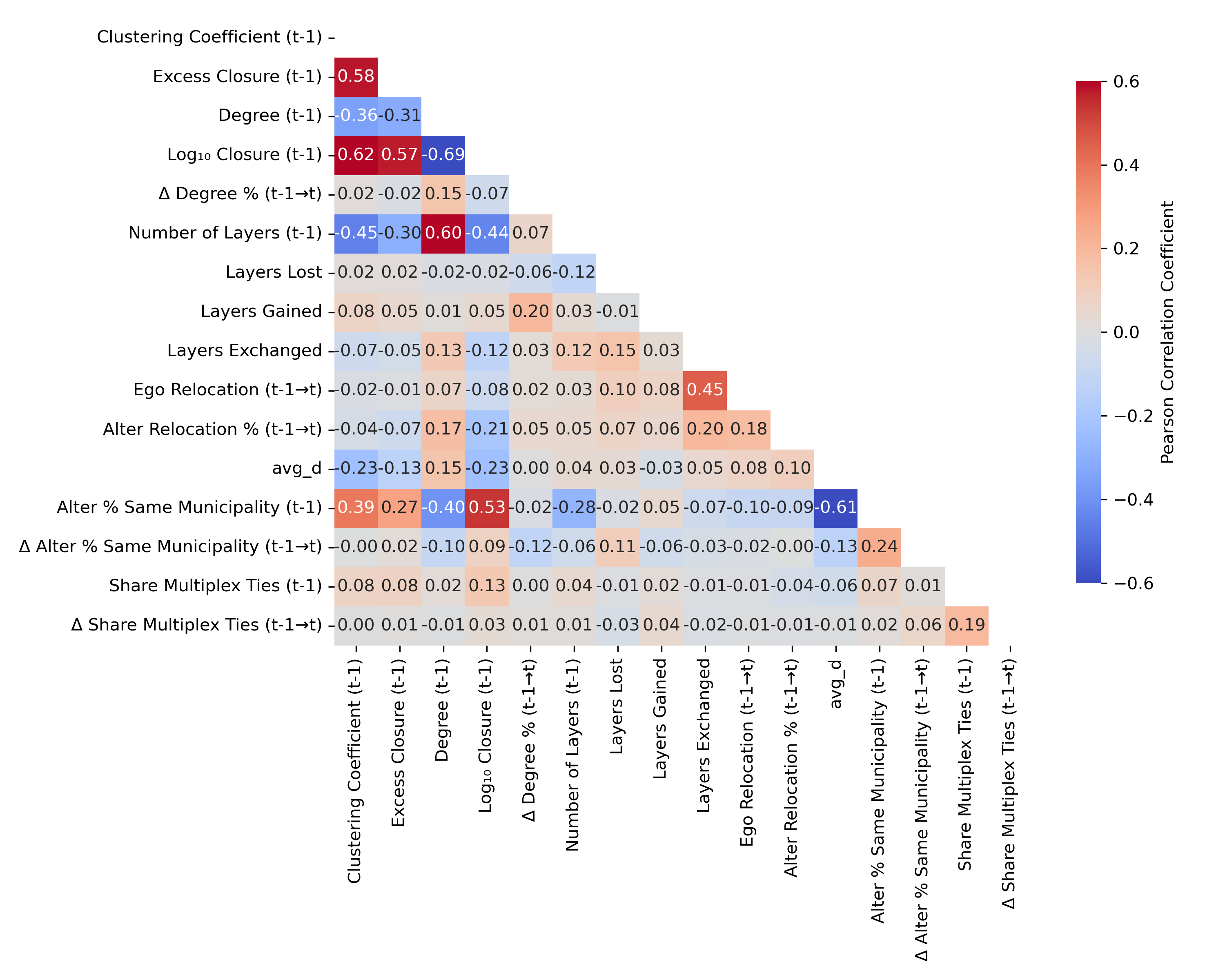}
    \caption{\textbf{Pearson correlation matrix of network, spatial, and demographic variables used in regression models.}
The heatmap displays pairwise correlations for 15 variables measured across 50,000 individuals over 
12 years in the panel database (2010--2021).}
    \label{fig:corrplot}
\end{figure}

\newpage

\newgeometry{top=0.5cm,bottom=0.5cm}
\begin{landscape}
\begin{table}[t]
\centering
\caption{Global Properties}
\label{tab:global_properties}
\begin{tabular}{lrrrrrrr}
\toprule
Year & Nodes Total & Edges Family & Edges Household & Edges Neighbor & Edges School & Edges Work & Edges Total \\
\midrule
2010 & 16,577,672 & 155,509,162 & 21,455,123 & 238,510,720 & 147,238,136 & 414,808,666 & 958,404,720 \\
2011 & 16,655,963 & 156,548,191 & 22,448,441 & 238,840,886 & 145,213,754 & 418,863,120 & 962,744,740 \\
2012 & 16,730,274 & 157,394,292 & 21,542,455 & 238,940,302 & 139,484,046 & 413,476,568 & 951,789,839 \\
2013 & 16,779,733 & 158,045,663 & 20,807,206 & 238,081,405 & 135,671,408 & 406,478,640 & 940,274,476 \\
2014 & 16,828,663 & 158,694,584 & 20,550,583 & 237,640,473 & 133,432,741 & 406,531,363 & 938,222,813 \\
2015 & 16,900,111 & 159,359,397 & 20,539,936 & 237,702,578 & 129,749,321 & 412,384,638 & 941,187,239 \\
2016 & 16,978,518 & 159,854,778 & 20,624,944 & 237,908,701 & 126,918,371 & 418,670,220 & 945,439,977 \\
2017 & 17,080,849 & 160,419,995 & 21,663,411 & 239,114,217 & 125,599,606 & 429,268,679 & 957,448,439 \\
2018 & 17,180,445 & 160,942,307 & 22,776,608 & 240,095,208 & 129,933,378 & 441,388,945 & 976,417,898 \\
2019 & 17,281,594 & 161,347,071 & 22,792,509 & 240,725,031 & 135,124,979 & 448,798,950 & 990,035,615 \\
2020 & 17,407,004 & 161,753,829 & 22,054,674 & 241,642,746 & 141,162,126 & 442,081,800 & 990,007,126 \\
2021 & 17,474,700 & 162,467,827 & 19,887,776 & 241,341,724 & 149,555,858 & 452,661,842 & 1,007,195,644 \\
\bottomrule
\end{tabular}

\end{table}
\end{landscape}
\restoregeometry

\newpage

\newgeometry{top=0.5cm,bottom=0.5cm}
\begin{landscape}
\begin{table}[h!]
\caption{\textbf{Descriptive statistics of core network metrics across years (2010--2021).}}
\label{tab:si_core_metrics}
\centering
\small
\begin{tabular}{@{}llrrrrrrrrrrrr@{}}
\toprule
\textbf{Metric} & \textbf{Statistic} & \textbf{2010} & \textbf{2011} & \textbf{2012} & \textbf{2013} & \textbf{2014} & \textbf{2015} & \textbf{2016} & \textbf{2017} & \textbf{2018} & \textbf{2019} & \textbf{2020} & \textbf{2021} \\
\midrule
\multirow[c]{7}{*}{\begin{tabular}[c]{@{}l@{}}Clustering\\Coefficient\end{tabular}} 
 & Mean & 0.435 & 0.417 & 0.418 & 0.418 & 0.416 & 0.412 & 0.410 & 0.408 & 0.406 & 0.404 & 0.407 & 0.404 \\
 & Std & 0.156 & 0.160 & 0.159 & 0.159 & 0.157 & 0.156 & 0.157 & 0.157 & 0.157 & 0.155 & 0.155 & 0.156 \\
 & 10th & 0.276 & 0.251 & 0.253 & 0.255 & 0.253 & 0.249 & 0.245 & 0.244 & 0.243 & 0.243 & 0.247 & 0.243 \\
 & 25th & 0.336 & 0.313 & 0.314 & 0.316 & 0.313 & 0.310 & 0.306 & 0.305 & 0.302 & 0.302 & 0.305 & 0.301 \\
 & Median & 0.407 & 0.388 & 0.389 & 0.389 & 0.387 & 0.384 & 0.382 & 0.380 & 0.378 & 0.376 & 0.378 & 0.376 \\
 & 75th & 0.502 & 0.484 & 0.485 & 0.485 & 0.481 & 0.478 & 0.476 & 0.475 & 0.472 & 0.470 & 0.471 & 0.469 \\
 & 90th & 0.631 & 0.614 & 0.614 & 0.615 & 0.611 & 0.607 & 0.604 & 0.604 & 0.601 & 0.599 & 0.600 & 0.598 \\
\addlinespace
\multirow[c]{7}{*}{\begin{tabular}[c]{@{}l@{}}Excess\\Closure\end{tabular}} 
 & Mean & 0.161 & 0.156 & 0.154 & 0.154 & 0.151 & 0.150 & 0.147 & 0.144 & 0.140 & 0.138 & 0.139 & 0.136 \\
 & Std & 0.195 & 0.309 & 0.304 & 0.316 & 0.300 & 0.311 & 0.297 & 0.267 & 0.249 & 0.261 & 0.289 & 0.251 \\
 & 10th & 0.015 & 0.014 & 0.014 & 0.014 & 0.014 & 0.014 & 0.013 & 0.013 & 0.013 & 0.013 & 0.013 & 0.012 \\
 & 25th & 0.031 & 0.029 & 0.029 & 0.029 & 0.028 & 0.028 & 0.027 & 0.027 & 0.026 & 0.025 & 0.025 & 0.025 \\
 & Median & 0.098 & 0.086 & 0.086 & 0.086 & 0.084 & 0.083 & 0.081 & 0.079 & 0.076 & 0.074 & 0.075 & 0.073 \\
 & 75th & 0.225 & 0.198 & 0.195 & 0.194 & 0.191 & 0.188 & 0.185 & 0.182 & 0.177 & 0.171 & 0.172 & 0.167 \\
 & 90th & 0.385 & 0.359 & 0.356 & 0.356 & 0.352 & 0.352 & 0.347 & 0.339 & 0.336 & 0.328 & 0.329 & 0.324 \\
\addlinespace
\multirow[c]{7}{*}{\begin{tabular}[c]{@{}l@{}}Total\\Degree\end{tabular}} 
 & Mean & 117.6 & 126.8 & 125.6 & 123.4 & 123.2 & 123.3 & 123.1 & 124.3 & 126.1 & 127.0 & 126.0 & 127.8 \\
 & Std & 109.8 & 108.8 & 99.9 & 93.8 & 93.9 & 93.2 & 91.3 & 93.3 & 97.0 & 97.6 & 99.8 & 102.1 \\
 & 10th & 33 & 35 & 35 & 35 & 34 & 34 & 34 & 34 & 34 & 35 & 35 & 35 \\
 & 25th & 48 & 52 & 52 & 52 & 51 & 51 & 51 & 52 & 52 & 53 & 52 & 52 \\
 & Median & 85 & 97 & 96 & 95 & 94 & 94 & 95 & 95 & 97 & 98 & 96 & 98 \\
 & 75th & 170 & 182 & 181 & 180 & 180 & 180 & 181 & 182 & 185 & 184 & 182 & 183 \\
 & 90th & 214 & 233 & 233 & 231 & 231 & 232 & 233 & 234 & 236 & 236 & 235 & 237 \\
\addlinespace
\multirow[c]{7}{*}{\begin{tabular}[c]{@{}l@{}}Log$_{10}$\\Closure\end{tabular}} 
 & Mean & $-1.080$ & $-1.116$ & $-1.118$ & $-1.118$ & $-1.125$ & $-1.133$ & $-1.140$ & $-1.149$ & $-1.163$ & $-1.173$ & $-1.171$ & $-1.181$ \\
 & Std & 0.550 & 0.542 & 0.541 & 0.538 & 0.538 & 0.542 & 0.542 & 0.542 & 0.545 & 0.547 & 0.547 & 0.547 \\
 & 10th & $-1.821$ & $-1.838$ & $-1.836$ & $-1.835$ & $-1.845$ & $-1.854$ & $-1.863$ & $-1.868$ & $-1.881$ & $-1.888$ & $-1.886$ & $-1.902$ \\
 & 25th & $-1.498$ & $-1.533$ & $-1.538$ & $-1.534$ & $-1.541$ & $-1.553$ & $-1.558$ & $-1.567$ & $-1.584$ & $-1.593$ & $-1.592$ & $-1.602$ \\
 & Median & $-1.005$ & $-1.062$ & $-1.060$ & $-1.064$ & $-1.071$ & $-1.079$ & $-1.089$ & $-1.099$ & $-1.117$ & $-1.126$ & $-1.121$ & $-1.135$ \\
 & 75th & $-0.646$ & $-0.703$ & $-0.709$ & $-0.711$ & $-0.719$ & $-0.724$ & $-0.732$ & $-0.739$ & $-0.751$ & $-0.766$ & $-0.764$ & $-0.775$ \\
 & 90th & $-0.414$ & $-0.444$ & $-0.447$ & $-0.448$ & $-0.453$ & $-0.453$ & $-0.459$ & $-0.469$ & $-0.473$ & $-0.483$ & $-0.482$ & $-0.489$ \\
\addlinespace
\multirow[c]{7}{*}{\begin{tabular}[c]{@{}l@{}}$\Delta$ Degree\\(\%)\end{tabular}} 
 & Mean & --- & 0.188 & 0.063 & 0.059 & 0.074 & 0.067 & 0.069 & 0.076 & 0.082 & 0.073 & 0.055 & 0.077 \\
 & Std & --- & 2.424 & 0.706 & 1.009 & 1.322 & 0.593 & 1.326 & 0.823 & 0.698 & 0.641 & 0.693 & 0.806 \\
 & 10th & --- & $-0.138$ & $-0.192$ & $-0.190$ & $-0.179$ & $-0.167$ & $-0.171$ & $-0.169$ & $-0.172$ & $-0.185$ & $-0.198$ & $-0.173$ \\
 & 25th & --- & $-0.005$ & $-0.048$ & $-0.049$ & $-0.045$ & $-0.044$ & $-0.046$ & $-0.046$ & $-0.045$ & $-0.048$ & $-0.051$ & $-0.045$ \\
 & Median & --- & 0.070 & 0.000 & 0.000 & 0.000 & 0.000 & 0.000 & 0.000 & 0.000 & 0.000 & 0.000 & 0.000 \\
 & 75th & --- & 0.205 & 0.045 & 0.041 & 0.044 & 0.048 & 0.048 & 0.052 & 0.054 & 0.049 & 0.045 & 0.050 \\
 & 90th & --- & 0.461 & 0.211 & 0.184 & 0.208 & 0.222 & 0.215 & 0.241 & 0.267 & 0.246 & 0.209 & 0.250 \\
\bottomrule
\end{tabular}
\end{table}

\clearpage

\begin{table}[h!]
\caption{Descriptive statistics of multilayer network dynamics across years (2010--2021).}
\label{tab:si_layer_dynamics}
\centering
\small
\begin{tabular}{@{}llrrrrrrrrrrrr@{}}
\toprule
\textbf{Metric} & \textbf{Statistic} & \textbf{2010} & \textbf{2011} & \textbf{2012} & \textbf{2013} & \textbf{2014} & \textbf{2015} & \textbf{2016} & \textbf{2017} & \textbf{2018} & \textbf{2019} & \textbf{2020} & \textbf{2021} \\
\midrule
\multirow[c]{7}{*}{\begin{tabular}[c]{@{}l@{}}Number of\\Layers\end{tabular}} 
 & Mean & 3.533 & 3.529 & 3.521 & 3.510 & 3.504 & 3.500 & 3.495 & 3.493 & 3.504 & 3.508 & 3.499 & 3.505 \\
 & Std & 0.827 & 0.827 & 0.830 & 0.829 & 0.828 & 0.830 & 0.837 & 0.842 & 0.840 & 0.839 & 0.843 & 0.851 \\
 & 10th & 3 & 3 & 3 & 3 & 2 & 2 & 2 & 2 & 2 & 2 & 2 & 2 \\
 & 25th & 3 & 3 & 3 & 3 & 3 & 3 & 3 & 3 & 3 & 3 & 3 & 3 \\
 & Median & 4 & 4 & 4 & 4 & 4 & 4 & 4 & 4 & 4 & 4 & 4 & 4 \\
 & 75th & 4 & 4 & 4 & 4 & 4 & 4 & 4 & 4 & 4 & 4 & 4 & 4 \\
 & 90th & 4 & 4 & 4 & 4 & 4 & 4 & 4 & 4 & 4 & 4 & 4 & 4 \\
\addlinespace
\multirow[c]{2}{*}{\begin{tabular}[c]{@{}l@{}}Layers Lost\end{tabular}} 
 & Mean & 0.000 & 0.042 & 0.044 & 0.047 & 0.044 & 0.040 & 0.042 & 0.040 & 0.041 & 0.039 & 0.042 & 0.041 \\
 & Std & 0.000 & 0.216 & 0.219 & 0.231 & 0.220 & 0.209 & 0.215 & 0.214 & 0.210 & 0.208 & 0.215 & 0.211 \\
\addlinespace
\multirow[c]{2}{*}{\begin{tabular}[c]{@{}l@{}}Layers Gained\end{tabular}} 
 & Mean & 0.000 & 0.073 & 0.074 & 0.070 & 0.071 & 0.077 & 0.077 & 0.078 & 0.081 & 0.081 & 0.081 & 0.083 \\
 & Std & 0.000 & 0.388 & 0.387 & 0.375 & 0.377 & 0.406 & 0.402 & 0.398 & 0.410 & 0.416 & 0.420 & 0.413 \\
\addlinespace
\multirow[c]{2}{*}{\begin{tabular}[c]{@{}l@{}}Layers\\Exchanged\end{tabular}} 
 & Mean & 0.000 & 0.089 & 0.093 & 0.090 & 0.090 & 0.089 & 0.094 & 0.096 & 0.104 & 0.101 & 0.096 & 0.102 \\
 & Std & 0.000 & 0.373 & 0.381 & 0.370 & 0.372 & 0.366 & 0.375 & 0.374 & 0.389 & 0.384 & 0.373 & 0.381 \\
\addlinespace
\multirow[c]{2}{*}{\begin{tabular}[c]{@{}l@{}}Share\\Multiplex Ties\end{tabular}} 
 & Mean & 0.00285 & 0.00279 & 0.00272 & 0.00262 & 0.00261 & 0.00255 & 0.00252 & 0.00246 & 0.00242 & 0.00241 & 0.00235 & 0.00225 \\
 & Std & 0.01187 & 0.01227 & 0.01154 & 0.01120 & 0.01165 & 0.01111 & 0.01138 & 0.01114 & 0.01142 & 0.01222 & 0.01225 & 0.00990 \\
\addlinespace
\multirow[c]{2}{*}{\begin{tabular}[c]{@{}l@{}}$\Delta$ Share\\Multiplex Ties\end{tabular}} 
 & Mean & --- & $-0.000099$ & $-0.000079$ & $-0.000097$ & $-0.000066$ & $-0.000078$ & $-0.000054$ & $-0.000096$ & $-0.000081$ & $-0.000073$ & $-0.000106$ & $-0.000061$ \\
 & Std & --- & 0.00664 & 0.00635 & 0.00573 & 0.00633 & 0.00589 & 0.00565 & 0.00592 & 0.00557 & 0.00608 & 0.00731 & 0.00652 \\
\bottomrule
\end{tabular}
\end{table}

\clearpage

\begin{table}[h!]
\caption{Descriptive statistics of spatial dispersion and relocation patterns across years (2010--2021).}
\label{tab:si_spatial_metrics}
\centering
\small
\begin{tabular}{@{}llrrrrrrrrrrrr@{}}
\toprule
\textbf{Metric} & \textbf{Statistic} & \textbf{2010} & \textbf{2011} & \textbf{2012} & \textbf{2013} & \textbf{2014} & \textbf{2015} & \textbf{2016} & \textbf{2017} & \textbf{2018} & \textbf{2019} & \textbf{2020} & \textbf{2021} \\
\midrule
\multirow[c]{7}{*}{\begin{tabular}[c]{@{}l@{}}Avg Alter\\Distance (km)\end{tabular}} 
 & Mean & 11.055 & 13.153 & 13.341 & 13.497 & 13.601 & 13.605 & 13.687 & 13.806 & 13.896 & 14.000 & 14.213 & 14.431 \\
 & Std & 12.672 & 14.090 & 14.260 & 14.353 & 14.306 & 14.082 & 14.057 & 14.138 & 14.220 & 14.209 & 14.327 & 14.512 \\
 & 10th & 0.802 & 0.976 & 1.015 & 1.043 & 1.076 & 1.104 & 1.113 & 1.148 & 1.188 & 1.243 & 1.261 & 1.331 \\
 & 25th & 2.844 & 3.633 & 3.721 & 3.792 & 3.896 & 3.960 & 3.979 & 4.004 & 4.107 & 4.190 & 4.284 & 4.418 \\
 & Median & 7.191 & 8.984 & 9.115 & 9.252 & 9.403 & 9.490 & 9.618 & 9.743 & 9.822 & 9.891 & 10.111 & 10.361 \\
 & 75th & 14.629 & 17.784 & 18.034 & 18.345 & 18.557 & 18.605 & 18.778 & 18.881 & 19.024 & 19.176 & 19.487 & 19.716 \\
 & 90th & 25.568 & 30.268 & 30.747 & 31.059 & 31.081 & 30.993 & 31.411 & 31.589 & 31.545 & 31.774 & 32.062 & 32.295 \\
\addlinespace
\multirow[c]{7}{*}{\begin{tabular}[c]{@{}l@{}}Alter \% Same\\Municipality\end{tabular}} 
 & Mean & 0.690 & 0.658 & 0.655 & 0.654 & 0.654 & 0.652 & 0.650 & 0.646 & 0.644 & 0.643 & 0.639 & 0.635 \\
 & Std & 0.245 & 0.247 & 0.248 & 0.248 & 0.246 & 0.246 & 0.247 & 0.248 & 0.248 & 0.247 & 0.248 & 0.248 \\
 & 10th & 0.306 & 0.289 & 0.284 & 0.284 & 0.283 & 0.285 & 0.282 & 0.279 & 0.277 & 0.275 & 0.273 & 0.270 \\
 & 25th & 0.509 & 0.463 & 0.460 & 0.457 & 0.458 & 0.458 & 0.455 & 0.449 & 0.446 & 0.448 & 0.443 & 0.435 \\
 & Median & 0.747 & 0.698 & 0.696 & 0.692 & 0.691 & 0.686 & 0.683 & 0.679 & 0.677 & 0.674 & 0.667 & 0.663 \\
 & 75th & 0.896 & 0.868 & 0.866 & 0.867 & 0.864 & 0.862 & 0.860 & 0.857 & 0.856 & 0.854 & 0.851 & 0.846 \\
 & 90th & 0.976 & 0.967 & 0.966 & 0.964 & 0.963 & 0.963 & 0.963 & 0.962 & 0.960 & 0.958 & 0.957 & 0.953 \\
\addlinespace
\multirow[c]{7}{*}{\begin{tabular}[c]{@{}l@{}}$\Delta$ Alter \%\\Same Munic.\end{tabular}} 
 & Mean & --- & $-0.034$ & $-0.004$ & $-0.003$ & $-0.002$ & $-0.004$ & $-0.004$ & $-0.006$ & $-0.004$ & $-0.002$ & $-0.006$ & $-0.006$ \\
 & Std & --- & 0.127 & 0.117 & 0.117 & 0.116 & 0.114 & 0.115 & 0.118 & 0.119 & 0.120 & 0.118 & 0.118 \\
 & 10th & --- & $-0.144$ & $-0.065$ & $-0.060$ & $-0.061$ & $-0.064$ & $-0.065$ & $-0.072$ & $-0.073$ & $-0.070$ & $-0.072$ & $-0.073$ \\
 & 25th & --- & $-0.068$ & $-0.018$ & $-0.017$ & $-0.017$ & $-0.017$ & $-0.018$ & $-0.019$ & $-0.018$ & $-0.018$ & $-0.019$ & $-0.019$ \\
 & Median & --- & $-0.021$ & 0.000 & 0.000 & 0.000 & 0.000 & 0.000 & 0.000 & 0.000 & 0.000 & 0.000 & 0.000 \\
 & 75th & --- & 0.001 & 0.013 & 0.013 & 0.014 & 0.014 & 0.014 & 0.015 & 0.017 & 0.017 & 0.015 & 0.015 \\
 & 90th & --- & 0.046 & 0.057 & 0.058 & 0.058 & 0.059 & 0.056 & 0.057 & 0.065 & 0.071 & 0.058 & 0.060 \\
\addlinespace
\multirow[c]{2}{*}{\begin{tabular}[c]{@{}l@{}}Ego Relocation\\(municipality)\end{tabular}} 
 & Mean & --- & 0.031 & 0.032 & 0.030 & 0.031 & 0.033 & 0.035 & 0.036 & 0.040 & 0.039 & 0.037 & 0.040 \\
 & Std & --- & 0.174 & 0.177 & 0.170 & 0.172 & 0.180 & 0.184 & 0.186 & 0.195 & 0.194 & 0.189 & 0.195 \\
\addlinespace
\multirow[c]{7}{*}{\begin{tabular}[c]{@{}l@{}}Alter Relocation \%\\(municipality)\end{tabular}} 
 & Mean & --- & 0.081 & 0.081 & 0.081 & 0.081 & 0.085 & 0.090 & 0.094 & 0.099 & 0.093 & 0.092 & 0.096 \\
 & Std & --- & 0.066 & 0.067 & 0.067 & 0.069 & 0.067 & 0.067 & 0.068 & 0.068 & 0.066 & 0.066 & 0.068 \\
 & 10th & --- & 0.016 & 0.016 & 0.016 & 0.016 & 0.019 & 0.022 & 0.024 & 0.027 & 0.025 & 0.024 & 0.025 \\
 & 25th & --- & 0.039 & 0.039 & 0.039 & 0.038 & 0.043 & 0.048 & 0.051 & 0.057 & 0.052 & 0.051 & 0.054 \\
 & Median & --- & 0.069 & 0.068 & 0.069 & 0.067 & 0.072 & 0.079 & 0.083 & 0.089 & 0.083 & 0.083 & 0.086 \\
 & 75th & --- & 0.107 & 0.105 & 0.106 & 0.105 & 0.111 & 0.117 & 0.122 & 0.128 & 0.121 & 0.119 & 0.125 \\
 & 90th & --- & 0.156 & 0.156 & 0.155 & 0.156 & 0.162 & 0.168 & 0.171 & 0.176 & 0.167 & 0.167 & 0.174 \\
\bottomrule
\end{tabular}
\end{table}

\begin{table}[p]
\begin{center}
\captionsetup{justification=centering} 
\small
\caption{\textbf{Annual sample sizes and panel sample sizes (2010--2021).}\\ Panel sample size reflects individuals observed in consecutive years (t and t+1); not applicable in 2010.}
\label{tab:si_sample_sizes}
\begin{tabular}{@{}lrrrrrrrrrrrr@{}}
	\toprule
 & 2010 & 2011 & 2012 & 2013 & 2014 & 2015 & 2016 & 2017 & 2018 & 2019 & 2020 & 2021 \\
\midrule
Sample Size & 31,389 & 31,555 & 31,682 & 31,753 & 31,858 & 32,050 & 32,144 & 32,331 & 32,453 & 32,610 & 32,860 & 32,957 \\
Panel Sample Size & --- & 30,919 & 31,053 & 31,142 & 31,239 & 31,335 & 31,458 & 31,626 & 31,761 & 31,890 & 32,066 & 32,268 \\
\bottomrule
\end{tabular}
\label{tab:sample_size}
\end{center}
\end{table}

\end{landscape}
\restoregeometry

\begin{figure}
    \centering
    \includegraphics[width=\linewidth]{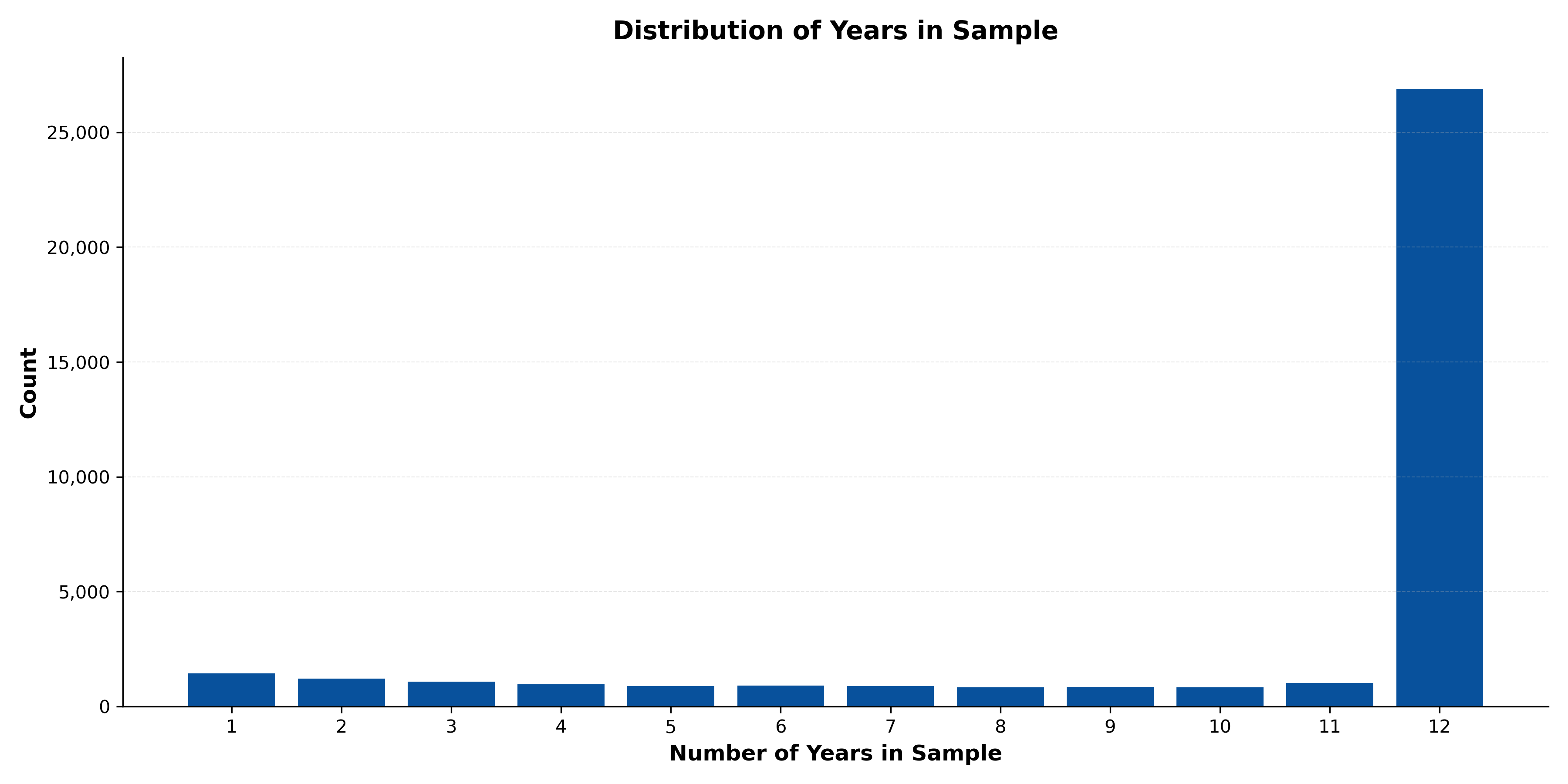}
    \caption{\textbf{Distribution of years in sample.} Histogram showing the count of individuals by the number 
of years they appear in the longitudinal sample (2010--2021). The majority of individuals ($N\approx 26,918$) 
are present for all 12 years.}
    \label{fig:years_in_sample}
\end{figure}

\begin{figure}[p]
    \centering
    \includegraphics[width=\linewidth]{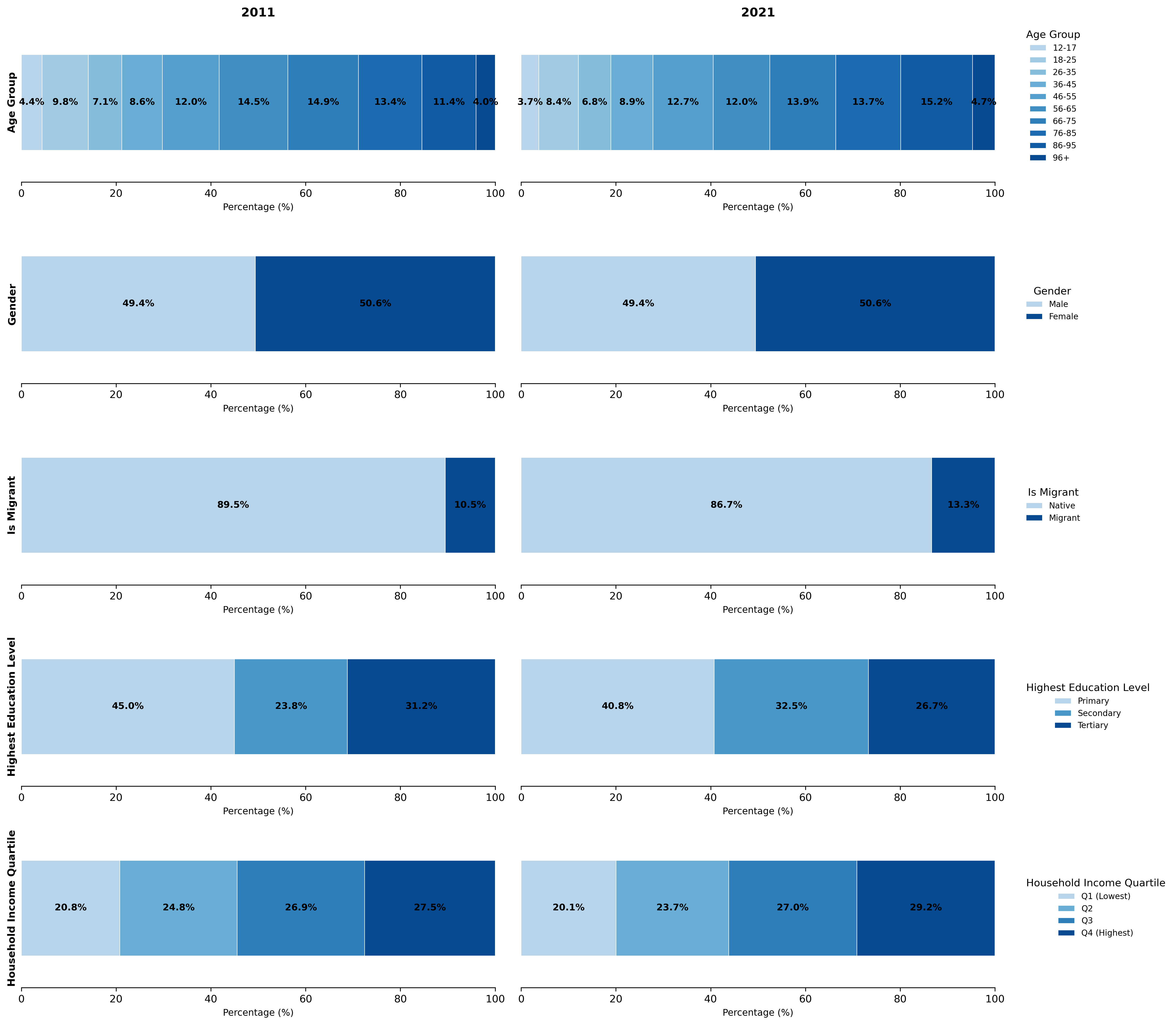}
    \caption{\textbf{Demographic composition of the sample in 2011 and 2021.} Each row represents a demographic 
variable: age group, gender, migration background, education level, and household income quartile. 
Horizontal stacked bars show the normalized percentage distribution of individuals across categories 
within each variable.}
    \label{fig:demogr_comp_2011_2021}
\end{figure}

\begin{figure}[p]
    \centering
    \includegraphics[width=\linewidth]{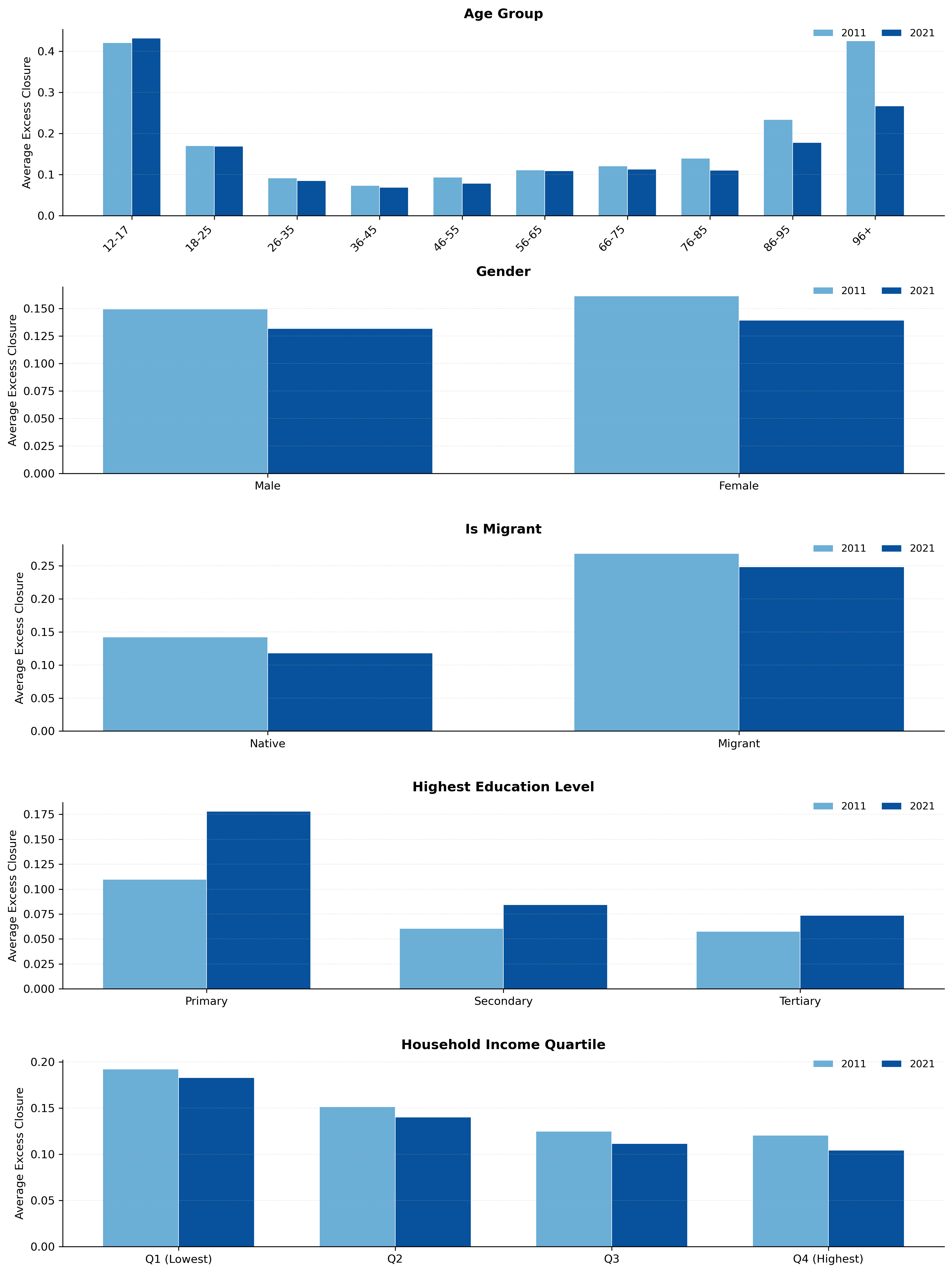}
    \caption{\textbf{Average excess closure by demographic group in 2011 and 2021.} Each panel shows the mean 
excess closure values for different categories within five demographic variables. Light blue bars 
represent 2011 values, dark blue bars represent 2021 values.}
    \label{fig:demogr_ec_2011_2021}
\end{figure}

\begin{figure}[p]   
    \centering
    \includegraphics[width=\linewidth]{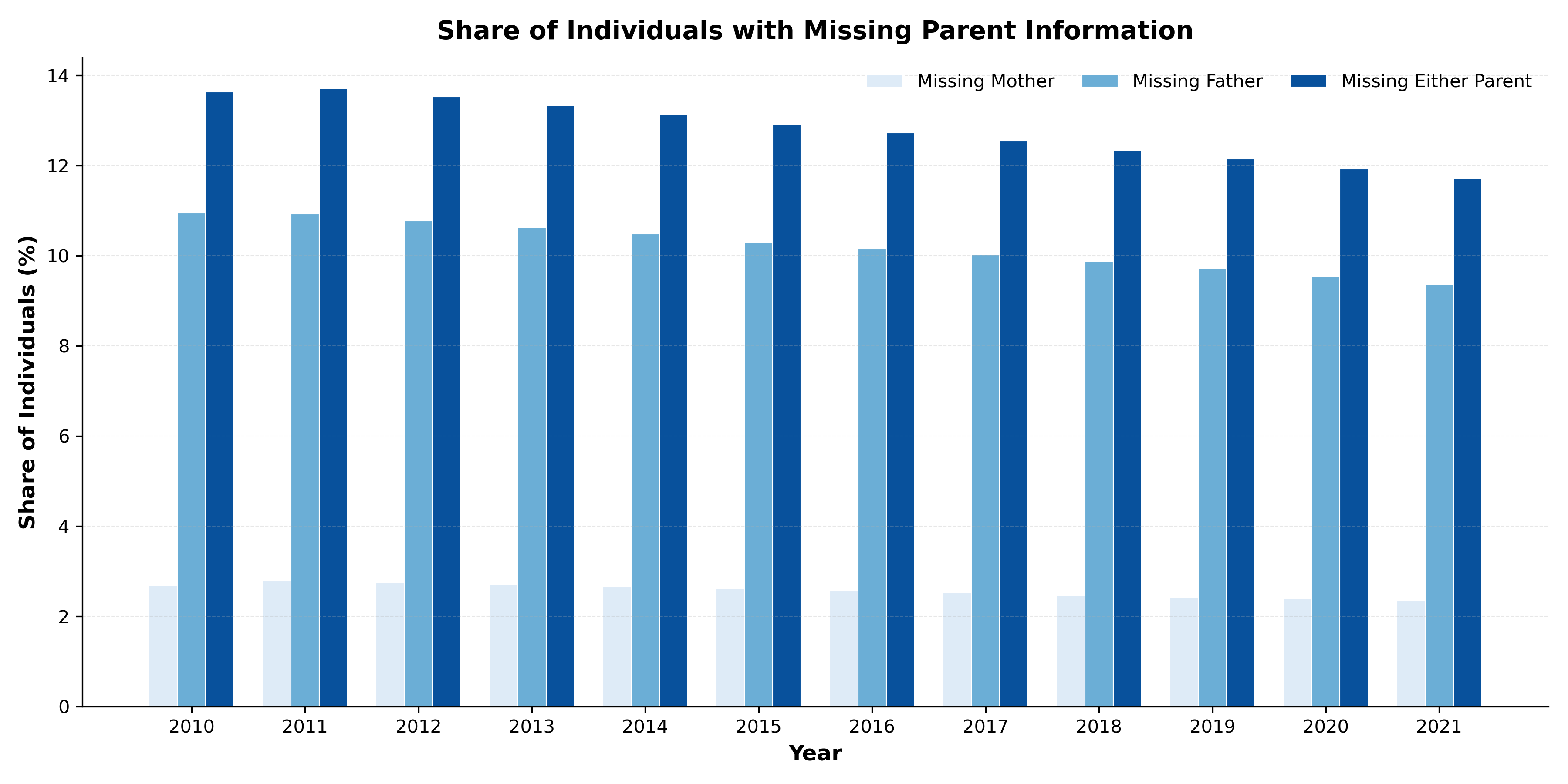}
    \caption{\textbf{Share of individuals with missing parent information by year (2010--2021)}. Grouped bars show 
the percentage of the sample with missing information for mother, father, or either parent across the 
12-year period. Light blue represents missing mother data, medium blue represents missing father data, 
and dark blue represents missing at least one parent.}
    \label{fig:share_missing_parents}
\end{figure}

\end{document}